\documentclass{svjour2}                    
\smartqed  
\usepackage{graphicx}
\usepackage{amssymb}
\usepackage{amsmath}
\usepackage{mathptmx}      
\usepackage{latexsym}
\newcommand{\rmd}{{\rm d}}

\newcommand{\CB}{{\cal B}}
\newcommand{\CD}{{\cal D}}
\newcommand{\CQ}{{Q}}
\newcommand{\CE}{{\cal E}}
\newcommand{\CW}{{\cal W}}
\newcommand{\CR}{{R}}
\newcommand{\CH}{{\cal H}}
\newcommand{\CM}{{\cal M}}
\newcommand{\CS}{{S}}
\newcommand{\CP}{{P}}

\newcommand{\averageM}[1]{\langle #1 \rangle_\CM}
\newcommand{\averageE}[1]{\langle #1 \rangle_\CE}
\newcommand{\average}[1]{\left\langle #1 \right\rangle_\CD}
\newcommand{\laverage}[1]{\left\langle #1 \right\rangle_{\CD_{\it i}}}
\newcommand{\gaverage}[1]{\left\langle #1 \right\rangle_{\Sigma}}
\newcommand{\initial}[1]{{#1_{\it i}}}
\newcommand{\now}[1]{{#1_{\it 0}}}
\newcommand{\inI}{{\rm I}}
\newcommand{\inII}{{\rm II}}
\newcommand{\inIII}{{\rm III}}
\journalname{Gen. Rel. Grav., Dark Energy Special Issue}
\begin{document}

\title{Dark Energy from structure: a status report} 
\titlerunning{Dark Energy from structure}        

\author{Thomas Buchert}

\institute{T. Buchert\qquad\;\; http://www.cosmunix.de\quad
              \email{buchert@cosmunix.de} \at
              \emph{Present address:} Laboratoire de l'Univers et ses Th\'eories LUTh, CNRS UMR 8102, 
Observatoire de Paris and Universit\'e Paris 7, 
F--92195 Meudon, France\\
             \emph{Permanent address:} Universit\'e Lyon 1, Centre de Recherche Astrophysique de Lyon, CNRS UMR 5574, 
9 avenue Charles Andr\'e, F--69230 Saint--Genis--Laval, France
}

\date{Received: 20 July 2007 / Accepted: 24 July 2007}

\maketitle

\begin{abstract}
The effective evolution of an inhomogeneous universe model in any theory of gravitation
may be described in terms of spatially averaged variables. In Einstein's theory, restricting
attention to scalar variables, this evolution can be modeled by solutions of a set of 
Friedmann equations for an effective volume scale factor, with matter and backreaction 
source terms. The latter can be represented by an effective scalar field  
(`morphon field') modeling Dark Energy.

The present work provides an overview over the Dark Energy debate in connection with the 
impact of inhomogeneities, and formulates strategies for a comprehensive quantitative evaluation 
of backreaction effects both in theoretical and observational cosmology. We recall the basic steps 
of a description of backreaction effects in relativistic cosmology that lead to refurnishing the 
standard cosmological equations, but also lay down a number of challenges and unresolved 
issues in connection with their observational interpretation.

The present status of this subject is intermediate: we have a good qualitative understanding of 
backreaction effects pointing to a global instability of the standard model of cosmology; 
exact solutions and perturbative results modeling this instability lie in the right sector to explain 
Dark Energy from inhomogeneities. It is fair to say that, even if backreaction effects turn out to be 
less important than anticipated by some researchers, the concordance high--precision cosmology, the
architecture of current N--body simulations, as well as standard perturbative approaches may all fall short 
in correctly  describing the Late Universe.
\end{abstract}
\section{General thoughts:\\$\quad$--- the standard model, the averaging problem and key insights}
\subsection{Views on and beyond the standard model of cosmology}
The standard model of cosmology does not, like  the standard model of particle physics, enjoy
appreciable generality; it is based on  
the simplest conceivable class of (homogeneous--isotropic) solutions of Einstein's laws of gravitation.
It is clear that the inhomogeneous properties of the Universe cannot be described by such a strong idealization.
The key issue is whether they can be described so {\it on average}, and this is the subject of considerable
debate and controversy in the recent literature. If the standard model indeed describes the
averaged model, we have to show that backreaction effects, being the main subject of this report, are negligible. 
We are striving to discuss most of the related aspects of this debate.

\subsubsection{Dark Energy and Dark Matter}

In the standard model of cosmology one has to conjecture the existence of two constituents,
if observational constraints are met, that both have yet unknown origin: 
first, a dominant repulsive component is thought to exist that
can be modeled either by a positive cosmological constant or a scalar field, e.g. a so--called 
quintessence field.
Besides this {\em Dark Energy},
there is, secondly, a non--baryonic component that should considerably exceed 
the contribution by luminous and dark baryons and 
massive neutrinos. This {\em Dark Matter} is thought to be provided by exotic forms of matter, 
not yet detected in (non--gravitational) experiments. According to the {\em concordance model} \cite{lahav}, \cite{SNLS}, \cite{spergeletal},
the former converges to about $3/4$ and the latter to about $1/4$ of the total source of
Friedmann's equations, up to a few percent that have to
be attributed to baryonic matter and neutrinos (in the matter--dominated era).
There are, however, other voices \cite{blanchardetal}, \cite{blanchard}.
 
Contemporary research to uncover this enigma pursues essentially two directions: one focusses on generalizations of the {\em geometry} of spacetime mostly restricting attention to modifications of the underlying theory of gravitation, the other invokes {\em new sources} in the energy momentum tensor and so implies a challenge for particle physics. As for the former, a Dark Energy component may possibly derive either from 
higher--order Ricci curvature Lagrangians \cite{capo} (as well as Capozziello and Francaviglia, this volume), \cite{riccilagrangians},
or string--motivated low--energy effective actions \cite{string:review}. 
It is doubtful whether a
{\em fundamental} scalar field exists in nature, at least one that can be viewed as a
natural candidate for the relevant effects needed to explain Dark Energy. This latter remark is supported by the well--known violation of energy conditions of a quintessence field that is able to produce late--time volume acceleration of the Universe.
Rather, a scalar field would likely be an {\em effective} one, either stemming from higher--order gravity terms, or effective terms as remnants from higher dimensions that are compactified or even
non--compactified as in brane world cosmologies \cite{maartens:brane} (see also Koyama, this volume).
As we shall learn below, already classical general relativity allows to
identify effective geometrical terms, simply resulting from inhomogeneities, with an effective scalar field component, the {\em morphon field} \cite{morphon}, a good example of  William of Ockham's razor. In this picture Dark Energy emerges as an excess of kinetic over potential energies of a scalar field in an `out--of--equilibrium' state, and it allows attributing Dark Energy to the classical vacuum. 
If we restrict our attention to cosmology and the fitting of extra terms from various different modified gravitational theories to observational data, then those extra terms may also be mapped into morphon fields with different but unambiguously defined physical consequences. 
A review of the status and properties of currently discussed models
can be found in  \cite{copeland:darkenergy}, see also \cite{paddy} (as well as Padmanabhan, this volume), \cite{uzan:darkenergy}, \cite{pilar}.
We shall not directly address the {\em Dark Matter problem} in this report, but also this problem might be related
to an explanation of Dark Energy; we shall  discuss such possible relations. 

Thus, the intriguing question is whether an explanation of these {\em dark} components is (i) the task
of particle physicists, or (ii) an expression of the need to modify the laws of gravitation, 
or (iii) whether the cosmological model is built on oversimplified priors.
We are going to study this last possibility.

\subsubsection{The longstanding averaging problem}

Does an inhomogeneous model of the Universe evolve {\em on average} like
a homogeneous solution of Einstein`s or Newton's laws of gravitation? 
This question is not new, at least among relativists who think that the answer
is certainly, in general, {\em no}, not only in view of the nonlinearity of the theories mentioned \cite{ellis}. 
The problem was and still is the notion of {\em averaging} whose specification and
unambiguous definition turned out to be an endeavor of high magnitude, mainly 
because it is not straightforward to give a unique meaning to the averaging
of tensors, e.g., a given metric of spacetime. This problem seems to lie in the
backyard of relativists who, from time to time, add another effort towards 
a solution of this technical issue. On the other hand, the community of cosmologists
{\em should} locate exactly this research topic at the basis of their evolutionary models of
the Universe. 

Although there have been numerous exceptions to this ubiquitous ignorance of the
averaging problem in cosmology, e.g. \cite{shirokovfisher}, and
many efforts after George Ellis \cite{ellis} has brought the subject into the fore, 
\cite{futamase1}, \cite{bildhauer:backreaction1}, \cite{bildhauer:backreaction2}, \cite{kasai1},
\cite{carfora:RG},
\cite{futamase2}, \cite{zala97}, \cite{ellisstoeger}, \cite{russetal}, \cite{boersma}, \cite{tanimoto}\footnote{This is certainly an incomplete list --
more references may be found in these papers and, e.g., in \cite{ellisbuchert}.}, still, the cosmologist's thinking 
rests on the hegemony of the {\it standard model} despite the drastic changes of our picture
of structures in the Universe on large scales.
This standard model, up to the present state of knowledge, is used as a prior to interpret a wide variety of
orthogonal observations, and it is therefore hard to beat due to this intentionally established status.
Therefore, most investigations in cosmology are still based on the 
vocabulary of the standard model, aiming to constrain its global {\it cosmological 
parameters}, often on the basis of observations of structure in the regional
Universe that is very different from homogeneous and isotropic.
As a consequence, also {\em structure} on large scales is described in terms of (quasi--Newtonian) perturbations
of this standard model, a construction that again makes only sense, if the standard model correctly describes
the average distributions of matter and geometry.     
Promisingly, the {\em conjecture} that the standard model agrees with the averaged model 
has recently been recognized as such and challenged by a wider community 
thanks to the Dark Energy debate.

\subsubsection{Uncharted territory beyond the standard model}

The {\em concordance model} is encircled by a large set of observational data that are,
however, orthogonal only within the predefined solution space of a FLRW 
(Friedmann--Lema\^\i tre--Robertson--Walker) cosmology. This solution space has dimension {\em two}
for Friedmann's expansion law derives from the Hamiltonian constraint of general relativity (see Eq.~(\ref{hamiltonlocal}) below), 
restricted to (about every point) locally
isotropic and hence (by Schur's Lemma) homogeneous distributions of matter and
curvature,
\begin{equation}
\label{hamilton1}
\Omega_m + \Omega_k + \Omega_{\Lambda}\;=\;1\;\;,
\end{equation}
where the standard cosmological parameters are {\em global} and iconized by the
{\em cosmic triangle} \cite{bahcall},
\begin{equation}
\label{cosmictriangle}
\Omega_m := \frac{8\pi G\varrho_H}{3 H^2}\;\;;\;\; \Omega_k := \frac{-k}{a^2 H^2}\;\;;\;\; \Omega_{\Lambda}:= \frac{\Lambda}{3 H^2}\;\;;
\end{equation} 
$\varrho_H (t)$ is the homogeneous matter density, $H(t):={\dot a} / a$ Hubble's function with the
scale factor $a(t)$, $k$ a positive, negative or
vanishing constant related to the three elementary constant--curvature geometries, and
$\Lambda$ is the cosmological constant, nowadays -- if positive -- employed as the simplest model of Dark Energy \cite{Peebles2003}.

We shall learn below that an extended solution space of an  averaged inhomogeneous universe model is three--dimensional, when we include inhomogeneities of matter and geometry. Hence, such more realistic models seem to enjoy more parameter freedom, but it should be 
emphasized that these (effective) `parameters' are defined in terms of volume averages of dynamically interacting physical variables. For a given inhomogeneous model, the additional parametrization appears in the initial conditions for the inhomogeneities that are absent
in the standard model of cosmology.
  
How can we be sure that fitting an idealized model, that ignores inhomogeneities, to observational data is not `epicyclic', especially if the model enters as a prior into the process of
interpreting the data?
Confronting observers with the wider class of averaged cosmologies allows them to draw their
data points within a cube of possible solutions {\em and} to differentiate the relevant observational scales reflected by these data; if we `force' them to draw the data points into the
plane of the FLRW solutions {\em on every scale}, then they conclude that there are  `dark' components. Thus, we have to exclude that they may have missed something in the projection {\em and} we have to clarify whether the ignorance of scale--dependence of observables in the standard model does not mislead their interpretation. 
Both issues are equally important to judge
the viability of the standard model in observational cosmology: the first is the question of
how backreaction {\em quantitatively} affects the standard cosmological parameters, and 
the second is the comparison of data taken on small scales (e.g. on cluster scales) and data
taken on large scales (e.g. CMB; high--redshift supernovae). Both additional `degrees of
freedom' in interpreting observational data are interlocked in the sense that backreaction
effects may alter the evolution history of cosmological parameters. A comparison of
data taken on different spatial scales has therefore also to be subjected to a critical assessment of data that are taken at different times of the cosmic history:
with backreaction at work, the simple time--scaling of parameters in a FLRW cosmology is also lost.

\vspace{3pt}

The plan of this report is the following. We shall first provide a list of arguments that justify existence of backreaction effects. Then, we move on to 
construct realistic universe models and discuss the governing equations in Section 2. A qualitative understanding of the backreaction mechanism relevant to the question of Dark Energy is developed in Section 3, and thereafter we propose and discuss strategies for a quantitative evaluation of backreaction effects in Section 4.  
Before we now enter the physics of backreaction that is easy to understand, we have to 
probe some more critical territory in the following subsection.

\subsection{Averaging strategies: different `directions' of backreaction}

The notion of {\em averaging} in cosmology is tied to space--plus--time thinking. Despite the success
of general covariance in the four--dimensional formulation of classical relativity, the
cosmologist's way of conceiving the Universe is {\em evolutionary}. This breaking of 
general covariance is in itself an obstacle to appreciating the proper status of cosmological
equations. The standard model of cosmology is employed with the implicit
understanding that there is a global {\em spatial} frame of reference that, if mapped to the
highly isotropic Cosmic Microwave Background, is elevated to a physical frame rather than
a particular choice of a mathematical slicing of spacetime. Restricting attention to an irrotational cosmic continuum of dust 
(that we shall retain throughout the main text), the best we can say is that all elements of the cosmic continuum defined by the
homogeneous distribution of matter are in {\em free fall} within that spacetime, and therefore
are preferred relative to accelerating observers with respect to this frame of
reference. Those preferred observers are called {\em fundamental}.
Exploiting the diffeomorphism degrees of freedom we can write the FLRW
cosmology in contrived ways, so that nobody would realize it as such. 
This point is raised as a criticism of an averaging framework \cite{wald}, as if this problem
were not there in the standard model of cosmology. Again, the `natural' choice for the matter model `irrotational dust' is a 
collection of {\em freely--falling} continuum elements, now for an inhomogeneous continuum.
For such a generalized collection of fundamental observers, the 4--metric form reads\footnote{For notations the reader may consult the 
Appendix; generally, we work with spatial variables in the hypersurfaces of constant coordinate time $t$ (that is equal to proper time for 
an irrotational dust continuum), and we explicitly indicate with a prefix when we
talk about four--dimensional variables in cases where this is not obvious.}
\begin{equation}
\label{fourmetric}
^4{\bf g} = -dt^2 + \;^3{\bf g}\;\;;\;\;^3{\bf g}=g_{ab}\,dX^a \otimes dX^b \;\;,
\end{equation}
where latin indices run through $1 \cdots 3$ and $X^a$ are local (Gaussian normal)
coordinates. Evolving the first fundamental form $^3{\bf g}$ of the spatial hypersurfaces
along $\partial / \partial t =:\partial_t$ defines their second fundamental form
\begin{equation}
\label{FF2}
^3{\bf K} = K_{ab}\,dX^a \otimes dX^b \;\;;\;\;K_{ab}:= -\frac{1}{2}\partial_t {g}_{ab}\;\;,
\end{equation}
with the extrinsic curvature components $K_{ab}$.
Such a {\em comoving (synchronous)} slicing of spacetime may be considered `natural', but it may also be
questioned. However, to dismiss its physical relevance due to the fact that shell--crossing
singularities arise is shortsighted. It is a problem of the {\em matter model} in the first place.
A comoving (Lagrangian) frame helps to access nonlinear stages of structure evolution,
as is well--exemplified in Newtonian models of structure formation, where the problem of choosing a proper slicing
is absent. Those nonlinear stages
inevitably include the development of singularities, provided we do not improve on the matter model
to include effects that counteract gravitation (like velocity dispersion) in order to regularize such 
singularities \cite{adhesive}. 
If a chosen slicing appears to be better suited, because it does not run into
singularities, then one should rather ask the question whether the evolution of variables
is restricted to a singularity--free regime just because inhomogeneities are not allowed to enter nonlinear stages of structure evolution. An example for this is perturbation theory formulated e.g. in longitudinal gauge, where the variables are `gauge--fixed' to a (up to a given time--dependent scale factor)
non--evolving background. 

However, the problem of choosing an appropriate slicing of spacetime is not off the table. There exist strategies to consolidate the notion
of an effective spatial slicing that would minimize frame fluctuations being attributed
to the diffeomorphism degrees of freedom in an inhomogeneous model. Such, more
involved, strategies 
relate to the  {\em intrinsic direction} of backreaction that we put into perspective below.

\subsubsection{Extrinsic (kinematical) and intrinsic backreaction}

Having chosen a {\em foliation of spacetime} implies that we can speak of two `directions':
one being {\em extrinsic} in the direction of the extrinsic curvature $K_{ab}$ of the embedding of the hypersurface into spacetime (e.g. parametrized by time),
the other being {\em intrinsic} in the direction of the Ricci tensor $R_{ab}$ of the three-dimensional
spatial hypersurfaces parametrized by a scaling parameter (let it be the geodesic radius of a randomly placed geodesic ball). 
Consequently, we may speak of two `directions' of
backreaction: inhomogeneities in extrinsic curvature and in intrinsic curvature.
The former is of {\em kinematical nature}, since we may interpret the extrinsic
curvature actively through the expansion tensor 
$\Theta_{ab}:=-K_{ab}$, and introduce a split into its kinematical parts: 
$\Theta_{ab} = 1/3 g_{ab}\Theta + \sigma_{ab}$, with the rate of expansion 
$\Theta =\Theta^c_{\;c}$, the shear tensor $\sigma_{ab}$, and the rate of shear
$\sigma^2 :=1/2 \sigma_{ab}\sigma^{ab}$;
note that vorticity and acceleration are absent for dust in the present  flow--orthogonal foliation. The latter addresses the so--called {\em fitting problem} \cite{ellis}, \cite{ellisstoeger},
\cite{singh2}, i.e. the question whether
we could find an effective constant--curvature geometry that best replaces the inhomogeneous
hypersurface at a given time. An answer to this question has to deal with the problem of `averaging' the tensorial (spatial) geometry
for which several different strategies are conceivable. Some of those strategies do not distinguish between extrinsic and intrinsic averaging (e.g. \cite{zala97}, \cite{zala05}, \cite{coley2}, and other 
references in  \cite{ellisbuchert}). A comparison of such a more `synthetic' approach with a pure kinematical averaging that leaves the physical
properties of a spatial hypersurface untouched has been provided \cite{singh3} and helps to also formally understand the differences between both viewpoints.
 
One method has recently
obtained a strong position in the context of Perelman's work (e.g. \cite{perelman:entropy}, \cite{perelman:ricci}) on the Ricci--Hamilton flow
related to the recent proof of Poincar\'e's conjecture, and implied progress on Thurston's geometrization program \cite{anderson} to cut a Riemannian manifold into `nice pieces' of eight elementary geometries. This method we briefly sketch now.

\subsubsection{Renormalization of average characteristics: smoothing the geometry}
\label{subsubsect:ricciflow}

Employing the Ricci--Hamilton flow \cite{hamilton:ricciflow1}, \cite{hamilton:ricciflow2},
\cite{carforamarzuoli},
an `averaging' of geometry can be put into practice by a rescaling of the spatial metric tensor, much in the spirit
of a renormalization flow \cite{carfora:RG}. 
A general scaling flow is described by Petersen's equations \cite{petersen} that we may implement through
a 2+1 setting by evolving the boundary of a geodesic ball in a three--dimensional
cosmological hypersurface in radial directions, thus exploring the Riemannian manifold
passively. Upon linearizing the general scaling flow, e.g. in normal geodesic coordinates,
we obtain a scaling equation for the metric along radial directions;
up to tangential geometrical terms on the boundary we obtain \cite{klingon},
\begin{equation}
\frac{\partial}{\partial r}\,g_{ab}(r) - \frac{\partial}{\partial r}\,g_{ab}(r)\Big\vert_{r_0} = -2 {\CR}_{ab} (r_0 ) [r-r_0 ]\;,
\end{equation}
i.e. the metric scales in the direction of its Ricci tensor much in the same way as it is deformed in the direction of the extrinsic curvature by the Einstein flow.
If we now implement the active (geometrically Lagrangian) point of view of deforming
the metric by the same flow along a Lagrangian vector field $\partial / \partial r_0$ while holding
the geodesic radius $r_0$ fixed, we are able to smooth the metric in a controlled way. Depending
on our choice of normalization of the flow, we may preserve the mass content inside
the geodesic ball while smoothing the metric.
Such a {\em mass--preserving Ricci flow} transforms kinematical averages on given hypersurfaces from their values
in the inhomogeneous geometry (the actual space section) to their values on a constant--curvature geometry (the fitting template for the space section): they are
{\em renormalized} resulting in additional backreaction effects due to the difference of the
two volumes (the Riemannian volume of the actual space section and the constant--curvature
volume) -- the {\em volume effect}, and also {\em curvature backreaction} terms that involve averaged invariants
of the Ricci tensor. For details and references see \cite{klingon} and for small overviews 
\cite{buchertcarfora:PRL} and \cite{buchertcarfora3}.
In such a setting the role of lapse and shift functions (i.e. the choice of slicing, {\it cf.} Appendix) can also be controlled by employing the recent 
results of Perelman \cite{carforabuchert:perelman}.

We now come to some crucial points of understanding the physics behind backreaction.
In order not to think of any exotic mechanism, the historical use of the notion `model with backreaction' should simply be replaced by `more realistic model'.

\subsection{The origin of kinematical backreaction and the physics behind it}

Let us now concentrate on the question, {\em why} there must be backreaction at work,
restricting attention to {\em kinematical backreaction} as defined above. In doing so, 
we do not actively modify the physics, i.e. the metrical properties of spatial sections; we
merely look at general {\em integral properties} of the inhomogeneous spatial distributions of matter and geometry on a given scale.
After we have understood the reasons behind backreaction effects in general terms, i.e. without resorting
to restrictions of spatial symmetry or approximations of evolution models, the very question of their relevance is better defined.

\subsubsection{An incomplete message to particle physicists}

Employing Einstein's general theory of relativity to describe the evolution of
the Universe, we base our universe model on a relation between geometry and matter sources. A maximal reduction
of this theoretical fundament is to consider the simplest conceivable geometry. 
Without putting in doubt that it might be an oversimplification to assume a (about every point) locally isotropic
(and hence homogeneous) geometry, standard cosmology conjectures the existence of sources that would generate this simple geometry. As already remarked, the majority of these
sources have yet unknown physical origin. Obviously, particle physicists take
the demand for missing fundamental fields literally. But, as was emphasized above, 
the standard model has physical sense only, if a homogeneous--isotropic solution of
Einstein's equations also describes
the inhomogeneous Universe effectively, i.e. {\em on average}.
This is not obvious. The very fact that the distributions of matter and geometry are
inhomogeneous gives rise to backreaction terms; we shall restrict them to those  additional terms that influence the kinematics of the homogeneous--isotropic solutions.
These terms can be viewed to arise on the geometrical side of Einstein's equations, but they
may as well be put on the side of the sources. 

We start with a basic kinematical observation that lies at the heart of the backreaction problem.

\subsubsection{A key to the averaging problem: non--commutativity}

Let us define spatial averaging of a scalar field $\Psi$ on a {\em compact}\footnote{This is a strong assumption on smaller spatial scales 
in the case of the matter model `irrotational dust': as soon as singularities in the flow develop, the boundary of the domain then also
experiences singularities, i.e. a breaking of the boundary due to a splitting of the domain or due to a merging of domains. These latter processes
that alter the domain's topology may also occur in a smooth way, if the flow is regularized through generalizations of the matter model.}
domain $\CD$ with volume $V_\CD : = \vert \CD \vert$ through its Riemannian volume average
\begin{equation}
\label{averager}
\average{\Psi (X^i ,t)}: = 
\frac{1}{V_\CD}\int_\CD  \Psi (X^i, t) \,J d^3 X\;\;\;;\;\;\;J:=\sqrt{\det(g_{ij})}\;.
\end{equation}
The key property of inhomogeneity of the field 
$\Psi$ is revealed by the {\em commutation rule} \cite{buchertehlers}, \cite{buchert:grgdust}:
\begin{equation}
\label{commutationrule}
\partial_t \langle\Psi \rangle_\CD - 
\langle\partial_t \Psi \rangle_\CD 
= \average{\Theta\Psi} - \average{\Theta}\average{\Psi}\;,
\end{equation}
where $\Theta :=u^{\mu}_{\;\,;\mu}$ denotes the trace of the fluid's expansion tensor, $u^{\mu}$ its 4--velocity, and $\partial_t J = \Theta J$ the evolution
of the root of the 3--metric determinant $J$; the spatial average of $\Theta$ describes the rate of volume change of
a collection of fluid elements along $\partial /\partial t$,
\begin{equation}
\label{effectivehubble}
 \average{\Theta} = \frac{\partial_t V_\CD}{V_\CD} = : 3H_\CD \;,
\end{equation} 
where we have introduced a volume Hubble rate $H_\CD$ that reduces to Hubble's function
in the homogeneous case.
{\em Commutativity} reflects the conjecture implied by the standard model: a realistically evolved inhomogeneous field will feature the same average characteristics as those predicted by the evolution
of the (homogeneous) average quantity; in other words, the right--hand--side of 
(\ref{commutationrule}) is assumed to vanish. This rule also shows that backreaction terms
deal with the sources of {\em non--commutativity} that are in general non--zero for inhomogeneous fields. 
Note that this rule is purely {\em kinematical}, which shows that it is not necessarily the nonlinearity of the field equations that is
responsible for backreaction effects. 

\subsubsection{Regional volume acceleration despite local deceleration}

Based on a first application of the above rule, we shall 
emphasize that there is not necessarily {\em anti--gravity} at work, e.g. in
the `redcapped' version of a positive cosmological constant, in order to have sources
that counteract gravity. Raychaudhuri's equation, if physically essential terms like
vorticity, velocity dispersion, or pressure are retained, provides terms needed 
to oppose gravity, e.g., to support spiral galaxies (vorticity), elliptical galaxies (velocity
dispersion), and other stabilization mechanisms involving pressure (think of the hierarchy of stable states of stars
until they collapse into a Black Hole). Admittedly, those terms
are effectively `small--scale--players'. Now, let us consider Raychaudhuri's equation (see (\ref{einstein}) below), restricted
to irrotational dust\footnote{We assume that the influence of a strong vorticity evolution (that is known to happen on small scales in the nonlinear regime of structure formation) is not relevant on scales larger than the scale of, say, superclusters of galaxies. According to the sign of its appearence in
Raychaudhuri's equation, vorticity counteracts gravitation and its effect will be relevant, if averages are performed over domains on and below the scales of galaxy clusters.},
\begin{equation}
\label{localraychaudhuri}
\partial_t \Theta = \Lambda - 4\pi G \varrho + 2{\rm II} - {\rm I}^2 \;\;,
\end{equation}
with the principal scalar invariants of $\Theta_{ab}$, $2{\rm II}:= 2/3 \Theta^2 - 2\sigma^2$ and ${\rm I}:=\Theta$.
Then, unless there is a positive cosmological constant, there is no term that could 
counter--balance gravitational attraction and, at every point, ${\partial_t \Theta} <0$.
Applying the {\em commutation rule} ({\ref{commutationrule}) for $\Psi = \Theta$,
we find that the averaged variables obey the same equation as above {\em despite}
non--commutativity\footnote{This is
only true, if all terms appearing in Raychaudhuri's equation are written in terms of principal scalar invariants;
it is actually a special non--linearity of this equation that cancels the corresponding non--commutativity
term (see Corollary I in \cite{buchert:grgdust}).}:
\begin{equation}
\label{regionalraychaudhuri}
\partial_t \average{\Theta} = \Lambda - 4\pi G \average{\varrho} + 2\average{{\rm II}} - 
\average{\rm I}^2\;\;.  
\end{equation}
This result can be understood on the grounds that shrinking the domain $\CD$ to a point
should produce the corresponding local equation. Now, notwithstanding, the above equation
contains a positive term that acts against gravity. This can be easily seen by rewriting the
averaged principal invariants: we obtain\footnote{We have formally inserted the averaged shear term, so that the last two
terms correspond to the local ones.} 
\begin{equation}
2\average{\rm II} - \average{\rm I}^2 = 
\frac{2}{3} \average{(\Theta - \average{\Theta})^2} - 2 \average{(\sigma - \average{\sigma})^2} - \frac{1}{3} \average{\Theta}^2 - 2\average{\sigma}^2\;\;,
\end{equation}
which, compared with
the corresponding local expression, 
\begin{equation}
2{\rm II}- {\rm I}^2 = -\frac{1}{3} \Theta^2 -2\sigma^2\;\;,
\end{equation}
gave rise to two additional, positive--definite fluctuation terms, where that for the averaged expansion variance enters with a positive sign.
It may appear
`magic' that the time--derivative of a (on some spatial domain $\CD$) averaged expansion may be positive despite the fact that the time--derivative of the expansion {\em at all points} in $\CD$ is negative. 
As the above explicit calculation shows, this property does not furnish an argument against the possibility of volume acceleration \cite{wald}, but simply is due to the fact that
an average correlates the local contributions, and it is this correlation (or fluctuation) that
adds `kinematical pressure'.  The interesting point is that these additional terms
are `large--scale players', as we shall make more precise below\footnote{The physical and observational consequences
of the expansion fluctuation term have been thoroughly explained and illustrated by a toy model in the review paper \cite{rasanen}.}.

What we can learn from this simple exercise is that any {\em local} argument, e.g.
on the smallness of some perturbation {\em amplitude} at a given point, is not enough to exclude 
{\em regional} (`global') physical effects that arise from averaging inhomogeneities; even if deviations from the average are small, as measured for example today, the evolution of the average may be different from the evolution of a `background solution' in perturbation theory.  As we shall discuss more in detail in the course of this report, such correlation effects must not be subdominant compared to the magnitude of the local fields,
since they are related to the spatial variation of the local fields and, having said `spatial', it could
(and it will) imply a coupling to the geometry as a dynamical variable in Einstein gravitation. 
This latter remark will turn out very useful in understanding the potential relevance of backreaction
effects in relativistic cosmology.

\subsubsection{The production of information in the Universe}
\label{subsect:information}

The above considerations on effective expansion properties can be essentially traced
back to `non--commutativity' of averaging and time--evolution, lying at the root of
backreaction. (Note that additional `spatial' backreaction terms that have been discussed
in Subsect.~\ref{subsubsect:ricciflow} 
are also the result of a `non--commutativity', this time
between averaging and spatial rescaling -- see also \cite{ellisbuchert}.) 
The same reasoning underlies the following entropy argument. Applying the {\em commutation rule}
(\ref{commutationrule}) to the density field, $\Psi = \varrho$,
\begin{equation}
\label{commutationentropy}
\average{\partial_t \varrho}-\partial_t \average{\varrho}\; =\;
\frac{{\partial_t \,{\CS}\lbrace\varrho || \average{\varrho}\rbrace}}
{V_{\CD}}\;\;,
\end{equation}
we derive, as a source of non--commutativity, 
the (for positive--definite density) positive--definite Lyapunov functional (known as {\it Kullback--Leibler functional} in information theory; \cite{hosoya:infoentropy} and 
references therein):
\begin{equation}
\label{relativeentropy}
{\CS} \lbrace \varrho || \average{\varrho}\rbrace: \;=\; 
\int_{\CD} \varrho \ln \frac{\varrho}{\average{\varrho}}\,J d^3 X \;\;.
\end{equation}
This measure vanishes for Friedmannian cosmologies (`zero structure'). 
It attains some {\em positive} time--dependent value otherwise.
The source in (\ref{commutationentropy}) shows that relative entropy production and
volume evolution are competing:  commutativity can be reached, if the volume expansion
is faster than the production of information contained within the same volume.
 
In \cite{hosoya:infoentropy} the following conjecture was advanced:

\smallskip

{\em The relative information entropy of a dust matter model
${\CS} \lbrace \varrho || \gaverage{\varrho}\rbrace$
is, for sufficiently large times, globally (i.e. averaged over the whole
manifold $\Sigma$ that is assumed simply--connected and without boundary) an increasing function of time.}

\medskip

\noindent
This conjecture already holds for linearized scalar perturbations at a Friedmannian 
background (the growing--mode
solution of the linear theory of gravitational instability implies $\partial_t \,{\CS} >0$
and $\CS$ is, in general, time--convex, i.e. $\partial_t^2 \,{\CS} >0$). 
Generally, information entropy is produced, i.e. $\partial_t \,{\CS} >0$ with
\begin{equation}
\label{entropyproduction}
\frac{{\partial_t \,{\CS}\lbrace\varrho || \average{\varrho}\rbrace}}
{V_{\CD}} = -\average{\delta\varrho \Theta} = 
-\average{\varrho \delta\Theta} = -\average{\delta\varrho \delta\Theta}\;, 
\end{equation}
(and with the deviations of the local fields from their average values, e.g.  $\delta\varrho :=
\varrho - \average{\varrho}$), 
if the domain
$\CD$ contains more {\em expanding underdense} and {\em contracting overdense} regions than the opposite states 
{\em contracting underdense} and {\em expanding overdense} regions. 
The former states are clearly favoured in the course of evolution,
as can be seen in simulations of large--scale structure. 

There are essentially three lessons relevant to the origin of backreaction that can be learned here.
First, structure formation (or `information' contained in structures) installs a positive--definite
functional as a potential to increase the deviations from commutativity; it can therefore
not be statistically `averaged away' (the same remark applies to the averaged variance of the expansion rate discussed before). 
Second, gravitational instability acts in the form of a negative feedback that enhances 
structure (or `information'), i.e. it favours contracting clusters and expanding voids. This tendency 
is opposite to the thermodynamical interpretation within a closed system where such a 
relative entropy would decrease and the system would tend to thermodynamical equilibrium.
This is a result of the long--ranged nature of gravitation: the system contained within $\CD$ must
be treated as an open system.  Third, backreaction is a genuinely 
non--equilibrium phenomenon, thus, opening this subject also to the language of non--equilibrium thermodynamics \cite{prigogine}, \cite{schwarz}, \cite{zimdahl}, general questions of gravitational
entropy \cite{penrose1}, \cite{penrose2}, \cite{prokopec:entropy}, \cite{hosoya:infoentropy}, \cite{graventropy}, and observational measures using distances to equilibrium \cite{infoentropy:biesiada}. `Near--equilibrium' can only be maintained (not established) by a 
simultaneous strong volume expansion of the system. 
Later we discuss an example of a cosmos that is `out--of--equilibrium', i.e. settled in a state
far from a Friedmannian model that, this latter, can be associated with the relative equilibrium state 
${\CS} = 0$. 

In particular, we conclude that the standard model may be a good description for the averaged
variables only when information entropy production is {\em over}--compensated by volume expansion
(measured in terms of a corresponding adimensional quantity). This latter
property is realized by linear perturbations at a FLRW background. Thus, the question
is whether this remains true in the nonlinear regime, where information production is strongly promoted by structure formation and
expected to be more efficient.

\bigskip

Before we can go deeper into the problem of whether such backreaction terms, being
well--motivated, are indeed relevant in a {\em quantitative} sense, we have to study the governing
equations.

\newpage

\section{Constructing a realistic universe model:\\\quad--- refurnishing the cosmological equations}

In this section we recall a set of averaged Einstein equations together with alternative
forms of these equations which put us in the position to study backreaction terms as
additional sources to the standard Friedmann equations.

\subsection{Einstein's equations recalled}

In order to make the presentation more self--contained, we recall the complete
set of local Einstein equations,
restricted to irrotational fluid motion
with the simplest matter model `dust' (i.e. vanishing pressure), as before\footnote{The corresponding equations with arbitrary lapse and 
shift functions for a perfect fluid energy--momentum--tensor are discussed in the Appendix,
together with the averaged equations.}.
In this case the flow is geodesic and space--like hypersurfaces can be
constructed that are flow--orthogonal at every spacetime event in a $3+1$ 
representation.

\noindent
We start with Einstein's equations\footnote{Greek indices run through
$0 ... 3$, while latin indices run through $1 ... 3$; summation over 
repeated indices is understood. A semicolon will denote covariant derivative with 
respect to the 4--metric with signature $(-,+,+,+)$; the units are such that $c=1$;
further below,
a double vertical slash $_{||}$ denotes covariant 
derivative with respect to the $3-$metric
${g}_{ij}$, while a single vertical slash denotes partial derivative 
with respect to the local coordinates $X^i$;
The overdot denotes partial time--derivative (at constant $X^i$) as before,
here identical to the covariant time--derivative $\partial_t = u^{\mu}\partial_{\mu}$.}
\begin{equation}
\label{einstein1}
{}^4 R_{\mu\nu} - \frac{1}{2}g_{\mu\nu}{}^4 R = 8\pi G \varrho u_{\mu} u_{\nu}
-\Lambda g_{\mu\nu} \;\;, 
\end{equation}
with the 4--Ricci tensor ${}^4 R_{\mu\nu}$, its trace ${}^4 R$, the fluid's $4-$velocity 
$u^{\mu}$ ($u^{\mu}u_{\mu} = -1$), the cosmological constant $\Lambda$,
and the rest mass density $\varrho$ obeying the conservation law
\begin{equation}
\label{einstein2}
\left( \varrho u^{\mu} u^{\nu}\right)_{\;;\mu} \;=\;0\;\;.
\end{equation}
In a flow--orthogonal coordinate system
$x^{\mu} = (X^k ,t)$ (i.e., Gaussian or normal coordinates which are comoving
with the fluid) we can write $x^{\mu}=f^{\mu}(X^k ,t)$, and we have 
$u^{\mu}={\dot f}^{\mu} = (1,0,0,0)$ and 
$u_{\mu}={\dot f}_{\mu} = (-1,0,0,0)$.
These coordinates are defined such as to label
geodesics in spacetime, i.e., $u^{\nu}u^{\mu}_{\;\,;\nu}=0$. 

Defining the two fundamental forms as in Eqs.~(\ref{fourmetric}, \ref{FF2}), with the 3--metric
coefficients $g_{ij}$ and the extrinsic curvature coefficients $K_{ij}:=-h^{\mu}_{\;\,i}h^{\nu}_{\;\,j} 
u_{\mu;\nu}$ (projected into the
hypersurfaces orthogonal to $u_{\mu}$ with the help of  
$h_{\mu\nu}:= g_{\mu\nu} + u_{\mu}u_{\nu}$), 
Einstein's equations (\ref{einstein1}) together with  (\ref{einstein2})
(contracted with $u_{\nu}$) then are equivalent to the following system 
of equations \cite{adm1962}, \cite{smarryork},
consisting of the energy or {\em Hamiltonian constraint} and the momentum or {\em Codazzi constraints},
\begin{equation}
\label{hamiltonlocal}
\frac{1}{2}\left( {\CR} 
+ K^2 - K^i_{\,\;j} K^j_{\,\;i} \right) = 8\pi G\varrho + \Lambda \;\;\;;\;\;\;
K^i_{\,\;j || i} - K_{| j} = 0 \;\;,
\end{equation}
and the {\it evolution equations} for the density and the two fundamental
forms,
\begin{equation}
\partial_t \varrho =  K \varrho \;\;\;;\;\;\;
\partial_t {g}_{ij} = -2 \;\,{g}_{ik}K^k_{\,\;j} \;\;\;;\;\;\;
\partial_t {K}^i_{\,\;j} =  K K^i_{\,\;j} + {\CR}^i_{\,\;j} - 
(4\pi G \varrho + \Lambda)\delta^i_{\,\;j} \;.
\end{equation}
${\CR}:= {\CR}^i_{\,\;i}$ and $K:=K^i_{\,\;i}$ 
denote the traces of the spatial Ricci tensor  ${\CR}_{ij}$
and the extrinsic curvature $K_{ij}$, respectively. 
Expressing the latter in terms of kinematical quantities,
\begin{equation}
-K_{ij} = \Theta_{ij} = \sigma_{ij} + \frac{1}{3}\Theta g_{ij}\;\;\;;\;\;\;-K = \Theta 
\;\;,
\end{equation}
with the expansion $\Theta_{ij}$, the trace--free symmetric shear $\sigma_{ij}$,
and the rate of expansion $\Theta$, we may write the above equations in the form
\begin{eqnarray}
\frac{1}{2} {\CR}
+ \frac{1}{3}\Theta^2 - \sigma^2  = 8\pi G\varrho + \Lambda \;\;\;;\;\;\;
{\sigma}^i_{\,\;j || i} = \frac{2}{3} \Theta_{| j} \;\;;\nonumber\\
\partial_t \varrho = - \Theta \varrho \;\;;\;\;
\partial_t {g}_{ij} = 2 \;\,{g}_{ik}{\sigma}^k_{\,\;j} 
+ \frac{2}{3}\Theta {g}_{ik}{\delta}^k_{\,\;j}\;\;;\nonumber \\
\partial_t \Theta + \frac{1}{3}\Theta^2 + 2\sigma^2 + 4\pi G\varrho - 
\Lambda \;=\;0 \;\;;\nonumber\\
\partial_t {\sigma}^i_{\,\;j} + \Theta {\sigma}^i_{\,\;j} = -
\left({\CR}^i_{\,\;j} -\frac{1}{3}{\delta}^i_{\,\;j}{\CR}\right)\;\;,
\label{einstein}
\end{eqnarray}
where we have introduced the rate of shear $\sigma^2 := 1/2 \sigma^i_{\;\,j} 
\sigma^j_{\;\,i}$.
(To derive the last two equations, {\em Raychaudhuri's equation} \cite{raychaudhuri} and the equation for the trace--free parts,
we have used the Hamiltonian constraint.)

\subsection{Averaged cosmological equations} 

In order to find evolution equations for effective (i.e. spatially averaged) cosmological
variables, we may put the following simple idea into practice. We observe that 
Friedmann's differential equations \cite{friedmann1}, \cite{friedmann2} capture the scalar parts of Einstein's equations (\ref{einstein}), 
while restricting them by the strong symmetry assumption of local isotropy.
The resulting equations,
Friedmann's expansion law (the energy or Hamiltonian constraint)  and Friedmann's acceleration law (Raychaudhuri's equation), together with
restmass conservation,
\begin{equation}
\label{friedmann}
3\left(\frac{\dot a}{a}\right)^2 - 8\pi G\varrho_H  - \Lambda \;=\;-\frac{3k}{a^2}\;\;;\;\;
3\frac{\ddot a}{a} + 4\pi G\varrho_H - \Lambda\;=\;0\;\;;\;\;
\dot \varrho_H + 3\left(\frac{\dot a}{a}\right) \varrho_H = 0 \;\;,
\end{equation}
can be replaced by their spatially averaged, {\em general} counterparts
(for the details
the reader is referred to \cite{buchert:grgdust,buchert:grgfluid,buchert:static,morphon}):
\begin{equation}
\label{averagedhamilton}
3\left( \frac{{\dot a}_\CD}{a_\CD}\right)^2 - 8\pi G \average{\varrho}-\Lambda \;=\; - \frac{\average{\CR}+{\CQ}_\CD }{2} \;\;;
\end{equation}
\begin{equation}
\label{averagedraychaudhuri}
3\frac{{\ddot a}_\CD}{a_\CD} + 4\pi G \average{\varrho} -\Lambda\;=\; {\CQ}_\CD\;\;;
\end{equation}
\begin{equation}
\label{averagedcontinuity}
\langle{\varrho}\rangle\dot{}_\CD + 3\frac{{\dot a}_\CD}{a_\CD} \average{\varrho}=0\;\;.
\end{equation}
We have replaced the Friedmannian scale factor by the {\em volume scale factor} $a_\CD$, depending  on content,  shape and position of the domain of averaging $\CD$,  defined via the
domain's volume  $V_{\CD}(t)=|\CD|$, and the initial 
volume $V_{\initial\CD}=V_{\CD}(\initial{t})=|\initial{\CD}|$:
\begin{equation}
\label{volumescalefactor}
a_\CD (t) := \left( \frac{V_{\CD}(t)}{V_{\initial\CD}} \right)^{1/3} \;\;.
\end{equation}
Using a scale factor instead of the volume should not be confused with `isotropy'. The above
equations are general for the evolution of a mass--preserving, compact domain containing
an irrotational continuum of dust, i.e. they
provide a background--free and non--perturbative description of inhomogeneous {\em and} aniso\-tropic fields\footnote{One could, of course, introduce an isotropic or anisotropic  reference background  \cite{buchertehlers} or, explicitly isolate an averaged shear from the above
equations to study deviations
from the  kinematics of Bianchi--type models, as was done with some interesting results
in \cite{barrowtsagas}.}.   
The new term appearing in these equations, the {\em kinematical backreaction}, arises as
 a result of expansion and shear fluctuations:
\begin{equation}
\label{backreactionterm} 
{\CQ}_\CD : = 2 \average{\rm II} - \frac{2}{3}\average{\rm I}^2 =
\frac{2}{3}\average{\left(\theta - \average{\theta}\right)^2 } - 
2\average{\sigma^2} \;;
\end{equation}
$\rm I$  and $\rm II$  denote  the principal scalar invariants  of the  extrinsic
curvature, and the second equality  follows  by introducing the
decomposition
of the  extrinsic curvature into  the kinematical variables, as before. 
Also, it is not a surprise that the general averaged 3--Ricci curvature $\average{\CR}$ replaces the constant--curvature term in Friedmann's equations. 
Note also that the term $\CQ_\CD$ encoding the fluctuations has the particular structure of
vanishing
at a Friedmannian background, a property that it shares with gauge--invariant variables\footnote{In a quasi--Newtonian setting, where averages
are taken on the Euclidean or constant--curvature {\em background space}, the variable $\CQ_\CD$ 
is gauge--invariant to second--order in perturbation theory \cite{kolbetal}, \cite{lischwarz}, since 
this variable vanishes at the background \cite{stewartwalker}, \cite{stewart};
for related thoughts see \cite{singh3}, \cite{paranjape}.}. 

In the Friedmannian case, Eqs.~(\ref{friedmann}), the acceleration law arises as the time--derivative
of the expansion law, if the integrability condition of restmass conservation is respected,
i.e. the homogeneous density $\varrho_H \propto a^{-3}$. In the general case, however, 
restmass conservation is not sufficient. In addition to the (built--in) general integral of Eq.~(\ref{averagedcontinuity}), 
\begin{equation}
\label{averageddensity}
\average{\varrho} = \frac{\laverage{\varrho(\initial{t})}}{a_\CD^3} 
= \frac{M_\CD}{a_\CD^3 V_{\initial\CD}} \;\;;\;\;M_\CD = M_{\initial\CD}\;\;,
\end{equation}
we also have to respect the following {\em curvature--fluctuation--coupling}:
\begin{equation}
\label{integrability}
\frac{1}{a_\CD^6}\partial_t \left(\,{\CQ}_\CD \,a_\CD^6 \,\right) 
\;+\; \frac{1}{a_\CD^{2}} \;\partial_t \left(\,\average{\CR}a_\CD^2 \,
\right)\,=0\;.
\end{equation}
This relation will be key to understand how backreaction can take the role
of {\em Dark Energy}.

\subsection{Alternative forms of the averaged equations}

We here provide three compact forms of the averaged equations introduced above,
as well as some derived quantities.  They will prove useful for our
further discussion of the backreaction problem.

\subsubsection{Generalized expansion law}

The correspondence  between Friedmann's expansion law (the first equation in (\ref{friedmann})) and the
general expansion law (\ref{averagedhamilton}) can be made more explicit through formal integration of the integrability condition (\ref{integrability}):
\begin{equation}
\label{integrabilityintegral}
\frac{3k_{\initial\CD}}{a_\CD^2} - \frac{1}{ a_\CD^2} \int_{t_i}^t \,dt' \;
{\CQ}_\CD\; \frac{d}{dt'} a^2_\CD (t')
= \frac{1}{2}\left(\langle {\CR} \rangle_\CD + {\CQ}_\CD\right) \;\;. 
\end{equation}
The (domain--dependent) integration constant $k_{\initial\CD}$ relates the new terms to
the `constant--curvature part'. We insert this latter integral back into the expansion law
(\ref{averagedhamilton}) and obtain:
\begin{equation}
\label{generalexpansionlaw}
3\frac{\dot{a}_\CD^2 + k_{\initial\CD}}{a_\CD^2 } - 8\pi G \average{\varrho}
- \Lambda = \frac{1}{ a_\CD^2} \int_{\initial{t}}^t \rmd t'\ {\CQ}_\CD
\frac{\rmd }{\rmd t'} a^2_\CD(t')\;\;.
\end{equation}
This equation is formally equivalent to its Newtonian counterpart \cite{buchertehlers}.
It shows that, by eliminating the averaged scalar curvature, the whole history of the averaged 
kinematical fluctuations acts as a source of a generalized expansion law that features the
`Friedmannian part' on the left--hand--side of (\ref{generalexpansionlaw}).

\subsubsection{Effective Friedmannian framework}

We may also recast the general equations (\ref{averagedhamilton}, \ref{averagedraychaudhuri}, \ref{averagedcontinuity}, \ref{integrability}) 
by appealing to the Friedmannian framework.
This amounts to re--interpret geometrical terms, that arise through averaging, as 
effective {\em sources} within a Friedmannian setting.

In the present case the averaged equations may be written as standard zero--curvature
Friedmann equations for an {\em effective perfect fluid energy momentum tensor}
with new effective sources \cite{buchert:grgfluid}:
\begin{eqnarray}
\label{equationofstate}
\varrho^{\CD}_{\rm eff} = \average{\varrho}-\frac{1}{16\pi G}{\CQ}_\CD - 
\frac{1}{16\pi G}\average{\CR}\;\;;\nonumber\\
{p}^{\CD}_{\rm eff} =  -\frac{1}{16\pi G}{\CQ}_\CD + \frac{1}{48\pi
G}\average{\CR}\;\;\;.
\end{eqnarray}
\begin{eqnarray}
\label{effectivefriedmann}
3\left(\frac{{\dot a}_\CD}{a_\CD}\right)^2 - 8\pi G \varrho^{\CD}_{\rm eff} - \Lambda\;=\;0\;\;;\nonumber\\
3\frac{{\ddot a}_\CD}{a_\CD}  +  4\pi G (\varrho^{\CD}_{\rm eff}
+3{p}^{\CD}_{\rm eff}) - \Lambda \;=\;0\;\;;\nonumber\\
{\dot\varrho}^{\CD}_{\rm eff} + 
3 \frac{{\dot a}_\CD}{a_\CD} \left(\varrho^{\CD}_{\rm eff}
+{p}^{\CD}_{\rm eff} \right)\;=\;0\;\;.
\end{eqnarray}
Eqs.~(\ref{effectivefriedmann}) correspond to the equations
(\ref{averagedhamilton}),
(\ref{averagedraychaudhuri}), (\ref{averagedcontinuity}) and (\ref{integrability}), respectively. 

We notice that ${\CQ}_\CD$, if interpreted as a source, introduces a component with `stiff equation of state', $p^\CD_{\CQ} = \varrho^\CD_\CQ$, suggesting a correspondence with
a free scalar field (discussed in the next subsection), while the averaged scalar curvature introduces a component with `curvature equation of state'
$p^\CD_{\CR} = -1/3 \varrho^\CD_{\CR}$. Although we are dealing with dust matter, we appreciate a `geometrical pressure' in the 
effective energy--momentum tensor.

There is, of course, some ambiguity in defining the effective sources. We recall \cite{buchert:static}
that, firstly, it may
sometimes be useful to incorporate $\Lambda$ into the effective 
sources by defining $\varrho^\CD_{\rm eff\Lambda}: = \varrho^\CD_{\rm eff} + \Lambda/8\pi G$
and $p^\CD_{\rm eff\Lambda} := p^\CD_{\rm eff} - \Lambda/8\pi G$.
Secondly, we might add the `constant--curvature term' $3 k_{\initial\CD} / a_\CD^2$ to the expansion law in 
(\ref{effectivefriedmann}); if we wish to do
so, then the effective sources can be represented solely through the kinematical
backreaction term ${\CQ}_\CD$ and its time--integral. For this we have to exploit
the `Newtonian form', Eq.~(\ref{generalexpansionlaw}), and would have to define the 
effective sources as follows:
\begin{equation}
\label{equationofstate-k}
\hat{\varrho}^{\CD}_{\rm eff}: = \average{\varrho}+\frac{X_\CD}{16\pi G} 
\;\;;\;\;{\hat p}^{\CD}_{\rm eff}: =  -\frac{{\CQ}_\CD}{12\pi G} - \frac{X_\CD}{48\pi G}
\;\;;\;\;  X_\CD := \frac{2}{ a_\CD^2} \int_{\initial{t}}^t \rmd t'\ {\CQ}_\CD
\frac{\rmd }{\rmd t'} a^2_\CD(t')  \;.
\end{equation}
The integrated form of the integrability condition, Eq.~(\ref{integrabilityintegral}), then
allows to express $X_\CD$ again through the averaged scalar curvature, 
$X_\CD = 6k_{\initial\CD}/a_\CD^2 - {\CQ}_\CD - \average{\CR}$, and we obtain the sources 
corresponding to (\ref{equationofstate}), however, with a curvature source that 
captures the deviations $W_\CD = \average{\CR}-6 k_{\initial\CD}/a_\CD^2$ from a constant--curvature model:
\begin{equation}
\label{equationofstate-kk}
\hat{\varrho}^{\CD}_{\rm eff} = \average{\varrho}-\frac{{\CQ}_\CD}{16\pi G} - 
\frac{W_\CD }{16\pi G}
\;\;\;\;\;;\;\;\;\;\;{\hat p}^{\CD}_{\rm eff} =  -\frac{{\CQ}_\CD}{16\pi G} + 
\frac{W_\CD}{48\pi G}\;.
\end{equation}

\subsubsection{`Morphed' Friedmann cosmologies}
\label{subsect:morphon}

In the above--introduced framework we distinguish the averaged matter source on the one hand, and averaged 
sources due to geo\-metrical inhomogeneities stemming from extrinsic and intrinsic curvature
(kinematical backreaction terms) on the other. As shown above, 
the averaged equations can be written as standard Friedmann equations that are sourced 
by both. Thus, we have the choice to consider the averaged model as a (scale--dependent) `standard model' with matter source evolving in a {\em mean field} of backreaction terms.
This form of the equations is closest to the standard model of cosmology. It is a `morphed' 
Friedmann cosmology, sourced by matter and `morphed' by a (minimally coupled) scalar field, the {\em morphon field} \cite{morphon}. We write
(recall that we have no matter pressure source
here):
\begin{equation}
\label{morphon:sources}
\varrho^\CD_{\rm eff} =: \langle\varrho\rangle_{\cal D} +
\varrho^\CD_{\Phi}\;\;\;;\;\;\;
p^\CD_{\rm eff} =: p^\CD_{\Phi}\;\;,
\end{equation}
with 
\begin{equation}
\label{morphon:field}
\varrho^\CD_{\Phi}=\epsilon \frac{1}{2}{{\dot\Phi}_\CD}^2 + U_\CD\;\;\;;\;\;\;p^\CD_{\Phi} =
\epsilon \frac{1}{2}{{\dot\Phi}_\CD}^2 - U_\CD\;\;,
\end{equation}
where $\epsilon=+1$ for a standard scalar field (with positive kinetic energy), and 
$\epsilon=-1$ for a phantom scalar field (with negative kinetic energy)\footnote{We have chosen the letter $U$ for the potential
to avoid confusion with the volume functional; if $\epsilon$ is negative, a `ghost' can formally arise on the level of an
effective scalar field, although the underlying theory does not contain one.}.
Thus, in view of Eq.~(\ref{equationofstate}), we obtain the following correspondence:
\begin{equation}
\label{correspondence1}
-\frac{1}{8\pi G}{\CQ}_\CD \;=\; \epsilon {\dot\Phi}^2_\CD - U_\CD\;\;\;;\;\;\;
-\frac{1}{8\pi G}\average{\CR}= 3 U_\CD\;\;.
\end{equation} 
Inserting (\ref{correspondence1}) into the integrability condition (\ref{integrability}) 
then implies that $\Phi_\CD$, for 
${\dot\Phi}_\CD \ne 0$, obeys the (scale--dependent) Klein--Gordon equation\footnote{Note that the potential is not
restricted to depend only on $\Phi_\CD$ explicitly. An explicit dependence on the averaged density and on other variables of the system
(that can, however, be expressed in terms of these two variables) is generic.}:
\begin{equation}
\label{kleingordon}
{\ddot\Phi}_\CD + 3 H_{\cal D}{\dot\Phi}_\CD + 
\epsilon\frac{\partial}{\partial \Phi_\CD}U(\Phi_\CD , \average{\varrho})\;=\;0\;\;.
\end{equation} 
The above correspondence allows us to interpret the kinematical backreaction effects in terms
of properties of scalar field cosmologies, notably quintessence or phantom--quintessence
scenarii that are here routed back to models of inhomogeneities.
{\em Dark Energy emerges as unbalanced kinetic and potential energies due to structural inhomogeneities}\footnote{More precisely,
kinematical backreaction appears as excess of kinetic energy density over the `virial balance', {\it cf.} Eq.~(\ref{virialbalance}),
while the averaged scalar curvature of space sections is directly proportional to the potential energy density; e.g. a void (a `classical vacuum') with on average negative scalar curvature (a positive potential) can be attributed to a negative potential energy of a morphon field (`classical vacuum energy').}. 
For a full--scale discussion of this correspondence see \cite{morphon}.

\subsubsection{A note on closure assumptions}
\label{closure}

This system of the averaged equations in the various forms introduced above does not close unless we specify a model for the inhomogeneities.
Note that, if the system would close, this would mean that we solved the scalar parts of the
GR equations in
general by reducing them to a set of ordinary differential equations
on arbitrary scales. 
Closure assumptions have been studied by prescribing a {\it cosmic equation of state}
of the form 
$p^\CD_{\rm eff} = \beta (\varrho^\CD_{\rm eff}, a_{\cal D})$ \cite{buchert:darkenergy}, 
\cite{buchert:static}, or by prescribing the backreaction terms through {\em scaling solutions}, e.g.
${\CQ}_\CD \propto a_\CD^n$, parametrized by a scaling index $n$ \cite{morphon}.
We shall come back to the important question of how to close the averaged equations later
in Subsect.~\ref{subsect:approximations}.

\subsection{Derived dimensionless quantities}

For any quantitative discussion it is important to provide a set of dimensionless characteristics
that arise from the above framework.  

\subsubsection{The cosmic quartet}

We start by dividing the volume--averaged Hamiltonian constraint (\ref{averagedhamilton})
by the squared {\em volume Hubble functional} $H_\CD := {\dot a}_\CD / a_\CD$ introduced before.
Then, expressed through the following set of `parameters' \footnote{We shall, henceforth, call these characteristics `parameters', but the reader should keep in mind that these are  functionals on $\CD$. Moreover, they are dynamically coupled.},
\begin{equation}
\label{omega}
\Omega_m^{\CD} : = \frac{8\pi G}{3 H_{\CD}^2} \langle\varrho\rangle_{\cal D}  \;\;;\;\;
\Omega_{\Lambda}^{\CD} := \frac{\Lambda}{3 H_{\CD}^2 }\;\;;\;\;
\Omega_{\CR}^{\CD} := - \frac{\average{\CR}}{6 H_{\CD}^2 }\;\;;\;\;
\Omega_{\CQ}^{\CD} := - \frac{{\CQ}_{\CD}}{6 H_{\CD}^2 } \;\;,
\end{equation}
the averaged Hamiltonian constraint assumes the form of a {\it cosmic quartet} 
\cite{buchert:jgrg,buchertcarfora3}:
\begin{equation}
\label{hamiltonomega}
\Omega_m^{\CD}\;+\;\Omega_{\Lambda}^{\CD}\;+\;\Omega_{\CR}^{\CD}\;+\;
\Omega_{\CQ}^{\CD}\;=\;1\;\;,
\end{equation}
showing that the solution space of an averaged inhomogeneous cosmology is three--dimensional in the present framework.
In this set, the averaged scalar curvature parameter and the kinematical backreaction parameter
are directly expressed through $\average{\CR}$ and ${\CQ}_{\CD}$, respectively.
In order to compare this pair of parameters with the `constant--curvature parameter' 
that is the only curvature contribution
in standard cosmology to interpret
observational data, we can alternatively introduce the pair
\begin{equation}
\label{omeganewton}
\Omega_{k}^{\CD} := - \frac{k_{\initial\CD}}{a_\CD^2 H_{\CD}^2 }\;\;\;;\;\;\;
\Omega_{{\CQ}N}^{\CD} := \frac{1}{3 a_\CD^2 H_\CD^2}
\int_{\initial{t}}^t \rmd t'\ {\CQ}_\CD\frac{\rmd }{\rmd t'} a^2_\CD(t')\;\;,
\end{equation}
being related to the previous parameters by
\begin{equation}
\label{parameterrelation}
\Omega_{k}^{\CD} +\Omega_{{\CQ}N}^{\CD}\;=\; 
\Omega_{\CR}^{\CD} + \Omega_{\CQ}^{\CD}\;=:\Omega_X^\CD\;\;.
\end{equation}
After a little thought we see that both sides of this equality would mimick a Dark Energy component, $\Omega_X^\CD$,
in a Friedmannian model.
Note, in particular, that it is not the additional backreaction parameter alone that can play this role, but it is the joint action with the
(total) curvature parameter, or, looking to the left--hand--side, it is the cumulative effect acquired during the history of the backreaction parameter. A positive cosmological term would require this sum, or the effective history, respectively, to be positive.

\subsubsection{Volume state finders}

Like the volume scale factor $a_\CD$ and the volume Hubble rate $H_\CD$, we may 
introduce `parameters' for higher derivatives of the volume scale factor, e.g. the {\em volume deceleration}
\begin{equation}
\label{deceleration}
q^\CD := -\frac{{\ddot a}_\CD}{a_\CD}\frac{1}{H_\CD^2} = \frac{1}{2}
\Omega_m^{\CD} + 2 \Omega_{\CQ}^{\CD} - \Omega_{\Lambda}^{\CD}\;\;.
\end{equation}
Following  \cite{sahnietal,alametal} (see also \cite{evans} and  
references therein) we may also define the following
{\it volume state finders} involving the third derivative of the volume scale factor:
\begin{equation}
\label{statefinder1}
r^\CD := \frac{{\dddot a}_\CD}{a_\CD}\frac{1}{H_\CD^3} = 
\Omega_m^{\CD}(1 + 2 \Omega_{\CQ}^{\CD}) + 2 \Omega_{\CQ}^{\CD}
(1+4\Omega_{\CQ}^{\CD}) - \frac{2}{H_\CD}{\dot\Omega}_{\CQ}^{\CD} \;\;,
\end{equation}
and
\begin{equation}
\label{statefinder2}
s^\CD := \frac{r^\CD -1}{3(q^\CD - 1/2)}\;\;.
\end{equation} 
The above definitions are identical to those given in  \cite{sahnietal,alametal}, 
however, note the following obvious and subtle differences.
One of the obvious differences was already mentioned: while the usual state finders of 
a global homogeneous state in the 
standard model of cosmology are the same for every scale, the volume state finders defined
above are different for different scales. The other is the fact that the volume state finders 
apply to an inhomogeneous cosmology with arbitrary 3--metric, while the usual state finders
are restricted to a FLRW metric.  Besides these there is a more subtle difference, namely a degeneracy in the Dark Energy density parameter:
while \cite{sahnietal,alametal} denote (with obvious adaptation) 
$1-\Omega_m^\CD = \Omega_{X}^\CD$ we have from the Hamiltonian constraint 
(\ref{hamiltonomega}) $\Omega_{X}^\CD = \Omega_{\CQ}^{\CD} + \Omega_{\CR} ^\CD$, i.e. so--called $X$--matter (Dark Energy)
is composed of two physically distinct components.

\subsubsection{Cosmic equation of state and Dark Energy equation of state}

We already mentioned the possibility to characterize a solution of the averaged
equations by a {\em cosmic equation of state} 
$p^\CD_{\rm eff} = \beta (\varrho^\CD_{\rm eff}, a_{\cal D})$ with $w_{\rm eff}^\CD
:= p^\CD_{\rm eff}/\varrho^\CD_{\rm eff}$. Now, we may separately discuss (i.e. without matter source) the 
{\it morphon equation of state} that plays the role of the {\it Dark Energy
equation of state} \cite{morphon},
\begin{equation}
\label{EOS}
w^\CD_{\Phi} := \frac{{\CQ}_\CD - 1/3 \langle {\CR}\rangle_\CD}
{{\CQ}_\CD + \langle {\CR}\rangle_\CD}\;.
\end{equation}
We can express the volume state finders through this equation of state parameter and its first time--derivative:
\begin{equation}
\label{statefinder1w}
r^\CD = 1+ \frac{9}{2}w^\CD_{\Phi} ( 1+w^\CD_{\Phi}) (1- \Omega_m^\CD ) - \frac{3}{2}
\frac{{\dot w}^\CD_{\Phi}}{H_\CD} (1-\Omega_m^\CD )\;,
\end{equation}
and
\begin{equation}
\label{statefinder2w}
s^\CD =  1+w^\CD_{\Phi} -\frac{1}{3H_\CD}\frac{{\dot w}^\CD_{\Phi}}{w^\CD_{\Phi}}\;\;,
\end{equation}
being zero for $w^\CD_{\Phi} \equiv -1$, i.e. for the case of a (scale--dependent) cosmological constant.
As emphasized in \cite{sahnietal,alametal}, the above expressions have the advantage
that one can immediately infer the case of a constant Dark Energy equation of state,
so--called {\it quiessence models}, that here correspond to scaling solutions of the morphon field
with a constant fraction of kinetic to potential energies \cite{morphon}:
\begin{equation}
\label{quiessence}
\frac{2 E^\CD_{\rm kin}}{E^\CD_{\rm pot}} = \frac{\varepsilon{\dot\Phi}_\CD^2 V_\CD}{-U_\CD V_\CD} = 
-1 - \frac{3\CQ_\CD}{\average{\CR}} = 2\frac{w^\CD_{\Phi}+1}{w^\CD_{\Phi} -1}\;\;,
\end{equation}
where the case $\CQ_\CD =0$ (no kinematical backreaction), or $w^\CD_{\Phi} = -1/3$ (i.e. 
$\varrho^\CD_{\Phi} + 3p^\CD_{\Phi} = 0$) corresponds to the `virial condition' 
\begin{equation}
\label{virialbalance}
2\, E^\CD_{\rm kin} \,+\, E^\CD_{\rm pot} =\;0\;\;,
\end{equation} 
obeyed by the scale--dependent Friedmannian model\footnote{In the case of vanishing kinematical backreaction, the scalar field is present
for our definition of the correspondence and it models a constant--curvature term $\average{\CR} = 6 k_{\initial\CD}/a_\CD^2$. Alternatively, we could associate a morphon with the deviations $W_\CD$ from the constant--curvature model only.}. As has been already remarked, a non--vanishing backreaction
is associated with violation of `equilibrium'. Note that a morphon field does not violate
energy conditions as in the case of a fundamental scalar field, {\it cf.} Subsect.~\ref{subsubsect:accelerationcondition}.
Again it is worth emphasizing that the above--defined equations of state are scale--dependent.

With the help of these dimensionless parameters an inhomogeneous, anisotropic and scale--dependent state can be effectively characterized.

\section{Implications and further insights:\\\quad--- qualitative views on backreaction}

Having laid down a framework to characterize inhomogeneous cosmologies and having 
understood the physical nature of backreaction effects, does not entitle us to draw conclusions on
the {\em quantitative} importance of inhomogeneities for the global properties of world models.
It may well be that the robustness of the standard model also withstands this challenge.
A good example is provided by Newtonian cosmology that is our starting point for discussing
the implications of the present framework.

\subsection{Thoughts on Newtonian cosmology and N--body simulations}
\label{subsect:newtonian}

Analytical as well as numerical models for inhomogeneities are commonly studied within Newtonian cosmology. Essential cornerstones of our understanding of inhomogeneities
rest on the Euclidean notion of space and corresponding Euclidean spatial averages. 

\subsubsection{Global properties of Newtonian models}

The present framework can also be set up for the Newtonian equations and, indeed,
at the beginning of its development the main result on global properties of Newtonian
models was the confirmation of the FLRW cosmology as a correct model describing the averaged
inhomogeneous variables. Technically, this result is due to the fact that the averaged principal
invariants, encoded in $\CQ_\CD$, are complete divergences on Euclidean space sections and,
therefore, have to vanish on some scale where we impose periodic boundary conditions
on the deviation fields from the FLRW background.  The latter is a necessary requirement
to obtain unique solutions for Newtonian models (for details see \cite{buchertehlers}). 

This point is interesting in itself, because researchers who have set up cosmological
N--body simulations did not investigate backreaction: the vanishing of the averaged 
deviations from a FLRW background
is enforced by construction. The same remark applies to analytical models, where a
homogeneous background is introduced with the manifest implication of coinciding with
the averaged model, but without an explicit proof.
The outcome that a FLRW cosmology indeed describes the average of a general
Newtonian cosmology can be traced back to the (non--trivial) property that the
second principal invariant $\rm II$ appearing in $\CQ_\CD$ can indeed be written (like the first) as a complete divergence, {\it cf.} Eq.~(\ref{invariants})
below. 
Since this is not valid in Riemannian geometry, `global' backreaction effects -- if relevant -- entail the need of
generalizing current cosmological simulations and analytical models. If backreaction
is substantial, then current models must be considered as toy--models that have improved
our understanding of structure formation, but are inapplicable in circumstances where
the dynamics of geometry is a relevant issue. We shall learn that (i) these circumstances are
those needed to route Dark Energy back to inhomogeneities, and (ii) at the precision level 
at which currently cosmological parameters are determined, it can already be demonstrated that
backreaction might potentially be a non--negligible player in the Late Universe. 

While the last point will be touched upon in Section~4, there are a number of more points that
improve our qualitative understanding, to which we turn now.

\subsubsection{Morphological and statistical interpretation of backreaction}
\label{subsubsect:minkowski}

The expansion law, Eq.~(\ref{generalexpansionlaw}), is built on the rate of change
of a simple morphological quantity, the volume content of a domain. Although
functionally it depends on other morphological characteristics of a domain, it
does not explicitly provide information on their evolution. An evolution equation
for the backreaction term $\CQ_\CD$ is missing. This fact touches
on the problem of closing the hierarchy of dynamical evolution equations
mentioned in Subsect.~\ref{closure}.

We shall, in this subsection, provide a morphological interpretation of $\CQ_\CD$ 
that is possible in the Newtonian framework (the following considerations substantially
rely on the Euclidean geometry of space). This will improve our understanding of what
$\CQ_\CD$ actually measures, if geometry is not considered as a dynamical variable.
We know from previous remarks that the dynamical coupling of $\CQ_\CD$ to the
geometry of space sections will change this picture.

Let us focus our attention on the boundary of the spatial domain $\CD$. 
A priori, the location of this boundary in a non--evolving background space enjoys some freedom which we may
constrain by saying that the boundary coincides with a velocity front of the fluid
(hereby restricting attention to irrotational flows).
This way we employ the Legendrian point of view of velocity fronts that is dual
to the Lagrangian one of fluid trajectories. 
Let $S(x,y,z,t) = s(t)$ define a velocity front at Newtonian time $t$, ${\bf v}=
\nabla S$. 

Defining the unit normal vector $\bf n$ on the front, 
${\bf n} = \pm \nabla S / |\nabla S |$ (the sign depends on whether the domain is expanding or
collapsing),
the average expansion rate can be written as a flux integral using Gauss' theorem,
\begin{equation}
\average{\Theta} = \frac{1}{V_\CD}\int_{\CD} \nabla\cdot {\bf v} \;d^3 x = \frac{1}{V_\CD}
\int_{\partial\CD} \,{\bf v}\cdot{\bf {dS}} \;\;,
\end{equation}
with the Euclidean volume element $d^3 x$,
and the surface element $d\sigma$, ${\bf {dS}} = {\bf n}\, d\sigma$.  
We obtain the intuitive result that the 
average expansion rate is related to another morphological quantity of the domain,
the total area of the enclosing surface:
\begin{equation}
\average{\Theta} = \pm \frac{1}{V_\CD}\int_{\partial\CD} |\nabla S | \,d\sigma\;\;.
\end{equation}
The principal scalar invariants of the velocity gradient
$v_{i,j} =: S_{,ij}$
can be transformed into complete divergences of vector fields \cite{ehlersbuchert}:
\begin{eqnarray}
\label{invariants}
\inI(v_{i,j})  = \Theta = \nabla \cdot {\bf v}  \;\;\;;\;\;\;
\inII(v_{i,j}) = \omega^2 - \sigma^2 + \frac{1}{3} \Theta^2
 = \frac{1}{2}\nabla\cdot
\Big({\bf v} (\nabla\cdot{\bf v}) - ({\bf v}\cdot\nabla){\bf v} \Big) \;\;; \nonumber
\end{eqnarray}
\begin{eqnarray}
\inIII(v_{i,j}) = \frac{1}{9} \Theta^3 + 2\Theta (\sigma^2 + \frac{1}{3}\omega^2 ) + \sigma_{ij}\sigma_{jk}\sigma_{ki} - \sigma_{ij}\omega_i \omega_j
\qquad\nonumber\\
 = \frac{1}{3}\nabla\cdot\left( \frac{1}{2}\nabla\cdot
\Big( {\bf v}(\nabla\cdot{\bf v}) - ({\bf v}\cdot\nabla){\bf v} \Big) {\bf v} - 
\Big({\bf v}(\nabla\cdot{\bf v}) - ({\bf v}\cdot\nabla){\bf v} \Big)\cdot\nabla{\bf v} \right) \;\;. 
\end{eqnarray}
(With our assumptions $\omega$ in the above expressions vanishes identically.)\\
In obtaining these expressions, the flatness of space is used essentially.
Inserting the velocity potential and performing the spatial average, we obtain \cite{buchert:cup}:
\begin{eqnarray}
\average{\inII}=\frac{1}{V_\CD} \int_{\CD}  {\inII}\;d^3 x =\int_{\partial\CD} H\, |\nabla S |^2
d\sigma \;;\\
\average{\inIII} = \frac{1}{V_\CD} 
\int_\CD  {\inIII} \;d^3 x =\pm \int_{\partial\CD} G\,|\nabla S |^3  d\sigma \;\;,
\end{eqnarray}
where $H$ is the local mean curvature and $G$ 
the local Gaussian curvature at every point on the $2-$surface bounding the domain.
$|\nabla S | = \frac{ds}{dt}$ equals $1$, if the instrinsic arc--length $s$ of the
trajectories of fluid elements is used instead of the extrinsic Newtonian time $t$.
The averaged invariants comprise, together with the volume, a complete set of
morphological characteristics known as the
{\em Minkowski Functionals} $\CW_{\alpha}$ of a body:
\begin{eqnarray}
\CW_0 (s): = \int_{\CD} d^3 x = V_\CD \;\;\;\;;\;\;\;\;
\CW_1 (s): = \frac{1}{3}\int_{\partial\CD} d\sigma \;\;;\nonumber\\
\CW_2 (s): = \frac{1}{3}\int_{\partial\CD} H\;d\sigma \;\;\;\;;\;\;\;\;
\CW_3 (s): = \frac{1}{3}\int_{\partial\CD} G\;d\sigma =\frac{4\pi}{3}\chi\;\;.
\end{eqnarray}
The Euler--characteristic $\chi$ determines the topology of the domain and 
is assumed to be an integral of motion ($\chi = 1$), 
if the domain remains simply--connected\footnote{Notice that this may provide a morphological closure condition
for the hierarchy of evolution equations.}.

Thus, we have gained a morphological interpretation of the backreaction term: it 
can be entirely expressed through three of the four Minkowski Functionals:
\begin{equation}
\label{eq:backreaction-minkowski}
Q_\CD (s)= 6  \left(\frac{\CW_2}{\CW_0} - \frac{\CW_1^2}{\CW_0^2}\right) \;\;.
\end{equation}
The $\CW_{\alpha}\;;\;\alpha=0,1,2,3$ have been introduced
into cosmology in \cite{mecke:mf} 
in order to statistically assess morphological properties of cosmic structure.
Minkowski Functionals proved to be useful tools to also incorporate
information from higher--order correlations, e.g., in the distribution of galaxies,
galaxy clusters, density fields or cosmic microwave background temperature maps
(\cite{kerscher:mf}, \cite{kerscher:cl}, \cite{schmalzing:mf},
\cite{schmalzing:cmb}; see the review by Kerscher 
\cite{kerscher:review} and references therein). Related to the
morphology of individual domains is the study of 
building blocks of large--scale cosmic structure 
\cite{sahni:web}, \cite{schmalzing:web}.

For a ball with radius $R$ we have for the Minkowski Functionals:
\begin{equation}
\CW^{\CB_R}_0 (s): = \frac{4\pi}{3}R^3\;\;;\;\;
\CW^{\CB_R}_1 (s): = \frac{4\pi}{3}R^2 \;\;;\;\; 
\CW^{\CB_R}_2 (s): = \frac{4\pi}{3}R\;\;;\;\;
\CW^{\CB_R}_3 (s): = \frac{4\pi}{3}\;\;.
\end{equation}

Inserting these expressions into the backreaction term,
Eq.~(\ref{eq:backreaction-minkowski}), shows that $Q^{\CB_R}_\CD (s) = 0$, 
and we have proved
Newton`s `Iron Sphere Theorem', i.e. the fact that a spherically--symmetric configuration features the expansion properties of a homogeneous--isotropic
model\footnote{This can be shown explicitly by using a radially--symmetric velocity field \cite{bks}.}. Moreover, we can understand now that 
the backreaction term encodes the deviations of the domain`s morphology from 
that of a ball, a fact that we shall illustrate now with the 
help of Steiner`s formula of integral geometry (see also \cite{mecke:mf}).

Let $d\sigma^0$ be the surface element on the
unit sphere, then (according to the Gaussian map) 
$d\sigma = R_1 R_2 d\sigma^0$ is the surface element of a $2-$surface
with radii of curvature $R_1$ and $R_2$. Moving the surface a distance
$\varepsilon$ along its normal we get for the surface element of the parallel 
velocity front:
\begin{equation} 
\label{eq:steiner}
d\sigma^{\varepsilon} = (R_1 + \varepsilon) (R_2 + \varepsilon) d\sigma^0 = 
\frac{R_1 R_2 + \varepsilon (R_1 + R_2 ) + \varepsilon^2 }{R_1 R_2}d\sigma = 
(1 + \varepsilon 2  H + \varepsilon^2 G) d\sigma \,,
\end{equation}  
where 
\begin{equation}
H = \frac{1}{2}\left(\frac{1}{R_1} + \frac{1}{R_2}\right)\;\;\;;
\;\;\; G = \frac{1}{R_1 R_2} \;\;,
\end{equation}
are the mean curvature and Gaussian curvature of the front as before.

Integrating Eq.~(\ref{eq:steiner}) over the whole front we arrive at a relation 
between the total surface area $A_\CD$ of the front and $A_{\CD_\varepsilon}$ of its
parallel front. The gain in volume may then be expressed by an integral of
the resulting relation with respect
to $\varepsilon$ (which is known as {\em Steiner`s formula} defining the
Minkowski Functionals of a (convex) body in three spatial dimensions):
\begin{equation}
V_{\CD_\varepsilon} =  V_\CD + \int_0^{\varepsilon}\, d\varepsilon' A_{\CD_{\varepsilon'}} =
V_\CD + \varepsilon A_\CD + \varepsilon^2 \int_{\partial\CD} H\;d\sigma + 
\frac{1}{3} \varepsilon^3  \int_{\partial\CD} G\;d\sigma \;\;.
\end{equation}
An important lesson that can be learned here is that the backreaction term $\CQ_\CD$ obviously encodes {\em all}
orders of the N--point correlation functions, since the Minkow\-ski Funktionals have this property;
it is not merely a two--point term as the form of $\CQ_\CD$ as an averaged variance would suggest. 
In other words, a complete measurement of fluctuations must take into account that the domain is 
Lagrangian and the shape of the domain is an essential expression of the full N--point statistics
of the matter enclosed within $\CD$. (For further statistical considerations of backreaction 
in terms of given fluctuation spectra see \cite{bks}, \cite{abundance}).
Kinematically, Steiner's formula shows that the volume scale factor $a_\CD$, being defined through the volume in Eq.~(\ref{volumescalefactor}), also depends
on other morphological properties of $\CD$ in the course of evolution. In a comoving relativistic setting, the domain $\CD$ is frozen into the metric of spatial sections,
so that we also understand that an evolving geometry in general relativity takes the role of this shape--dependence in the Newtonian framework.   

\subsubsection{Backreaction views originating from Newtonian cosmology and relativistic
perturbation theory of a FLRW background}
\label{subsubsect:quasinewtonian}

We may place Newtonian models, but also relativistic models that suppress the coupling between fluctuations, encoded in $\CQ_\CD$,
and the geometry of space sections, into the same category:
as a rule of thumb we can say that any model that describes fluctuations on a Euclidean `background space' 
must be rejected as a potential candidate for a backreaction--driven cosmology. The reason is that fluctuations in those models can be subjected to
periodic boundary conditions implying a globally (on the periodicity scale) vanishing kinematical backreaction \cite{buchertehlers}.  The very
architecture of such models is simply too restrictive to account for a non--vanishing (Hubble--scale) $\CQ_\CD$ being a generic
property of relativistic models. Of course, also in those models, backreaction can be
investigated (a detailed investigation within Newtonian cosmology may be found in \cite{bks}
as well as an application on the abundance statistics of collapsed objects \cite{abundance}), but it is then only a rephrasing of the known {\em cosmic variance} within the standard model of cosmology. Nevertheless, the potential relevance of a non--vanishing backreaction can also be seen in Newtonian 
cosmology: in \cite{bks} it was found that the magnitude of  $\Omega_\CQ^\CD$ remains small throughout the evolution, being restricted to fall off to zero on some scale, but the indirect influence of a non--vanishing $\CQ_\CD$ in the interior of the periodic box is strongly seen in the other cosmological parameters. Thus,
independent of our statement of irrelevance of the magnitude of $\Omega_\CQ^\CD$ on large scales in Newtonian cosmology, backreaction is clearly an important player
to interpret cosmological parameters starting at scales of galaxy surveys, and it may here be a key to also understand the {\em Dark Matter problem},
{\it cf.} Subsect.~\ref{subsubsect:dark}.

We refer to the term `quasi--Newtonian' when we think of relativistic models that are
restricted to sit locally close to a Friedmannian state, as in standard gauge--invariant perturbation theory \cite{kodamasasaki}, \cite{mukhanovetal},
\cite{mukhanov}, their average
properties being evaluated on Euclidean space sections \cite{abramo}. 
Although we do not refer to the discussion of
structure on super--Hubble scales \cite{rasanen:darkenergy}, \cite{kolbetal:superhubble}, \cite{branden1}, \cite{parry}, the following 
consideration would also apply there. The
integrability condition (\ref{integrability}), in essence, spells out the generic 
coupling of kinematical fluctuations to the evolution of the averaged scalar curvature. Thus, the freedom taken by a generic model is carried by
a non--vanishing $\CQ_\CD$ (even if small) into changes of the other cosmological parameters, notably the averaged scalar curvature.
If that coupling is {\em absent} (even if $\CQ_\CD$ is non--zero), Eq.~(\ref{integrability}) shows that $\CQ_\CD \propto V_\CD^{-2}$ and $\average{\CR}\propto a_D^{-2}$, i.e. the averaged curvature evolves like a constant--curvature model, and backreaction decays more rapidly than the averaged density, $\average{\varrho}\propto V_\CD^{-1}$. In other words, backreaction cannot be
relevant today in all models that suppress this coupling (we shall make this more precise in the following).
Therefore, as another rule of thumb, we may say that {\em any
(relativistic) model that evolves curvature at or in the vicinity of the constant--curvature
model is rejected as a potential candidate for a backreaction--driven cosmology} \cite{morphon}.   

In summary, {\em Dark Energy cannot be
routed back to inhomogeneities on large scales in Newtonian and quasi--Newtonian models, but
a careful re--interpretation of cosmological parameters will have nevertheless to be envisaged}.

\subsection{Qualitative picture for backreaction--driven cosmologies}

Looking at the backreaction term $\CQ_\CD$, the relevant 
positive term that could potentially drive an accelerated expansion 
in accord with recent indications from supernovae data \cite{SNLS}, \cite{conleyetal}, \cite{spergeletal} (see also 
Leibundgut and Enqvist, this volume)\footnote{Note, however, that
the interpretation of volume acceleration in those data relies on the FLRW cosmology.
Backreaction could be influential and could change the interpretation of astronomical data 
also without featuring an accelerating phase.},
is the averaged variance of the rate of expansion, {\em cf.} Eq.~(\ref{backreactionterm}). 
This term, however, is quadratic and the averaging operation involves a division by the square of the volume.
How can we then expect that, in an expanding Universe, such a term can  
be of any relevance at the present time? Before we give an answer to this question, let us introduce a criterion
for a backreaction--driven cosmology that requires volume acceleration, i.e. we postulate high relevance of backreaction. 
This can be done with the help of the averaged equations
as has been advocated by Kolb {\em et al.} \cite{kolbetal,kolbetalc}.

\subsubsection{Acceleration and energy conditions}
\label{subsubsect:accelerationcondition}

Let us look at the general acceleration law (\ref{averagedraychaudhuri}),
and ask when we would find {\em volume acceleration} on a given patch of the spatial
hypersurface \cite{kolbetal}, \cite{buchert:darkenergy,buchert:static}:
\begin{equation}
\label{accelerationcondition}
3\frac{{\ddot a}_\CD}{a_\CD} = \Lambda 
-4\pi G \langle\varrho\rangle_\CD + {\CQ}_\CD \;>\;0\,.
\end{equation}
We find that, if there is no cosmological constant, the necessary condition ${\CQ}_\CD > 4\pi G\langle\varrho\rangle_\CD$  must be satisfied on a sufficiently large scale, at least at the present time.
This requires that ${\CQ}_\CD$ is positive, i.e. shear fluctuations are superseded by
expansion fluctuations\footnote{From the observational point of view this property is in accord with constraints that can be imposed on the
averaged shear fluctuations (quantitatively discussed in \cite{buchert:static}): 
the universe model can be highly isotropic in accord with strong constraints on the shear amplitude on large scales. For the backreaction term it is important to 
independently constrain the large--scale expansion fluctuations that are in general not necessarily proportional to large--scale density fluctuations as in a linear perturbation approach at a FLRW background.
Note also that the time--evolution of an isotropic average model must not (and in this case will not) coincide with the time--evolution of a FLRW background.}
{\em and}, what is crucial, that ${\CQ}_\CD$ decays less rapidly than the averaged density \cite{buchert:darkenergy}. 
It is not obvious that this latter condition could be met in view of our remarks
above. We conclude that backreaction has only a chance to be {\em relevant in magnitude compared with the density} 
(e.g. as defined through the inequality Eq.~(\ref{accelerationcondition}) today),
if its decay rate substantially deviates from its `quasi--Newtonian' behavior and, more precisely,
its decay rate must be weaker than that of the averaged density (or at least comparable,
depending on initial data for the magnitude of {\em Early Dark Energy}
\cite{caldwell&wetterich}, \cite{caldwell&linder}). 

Another model of Dark Energy is to assume the existence of a scalar field source, a so--called quintessence field (others are
discussed in \cite{copeland:darkenergy}). However, a usual scalar field source in a Friedmannian model, attributed 
e.g. to phantom quintessence that leads to acceleration, 
will violate the {\it strong energy condition} $\varrho + 3p >0$, i.e.:
\begin{equation}
\label{energyconditionF}
3\frac{\ddot a}{a} = -4\pi G (\varrho + 3p) = -4\pi G 
(\varrho_H + \varrho_{\Phi} + 3 p_{\Phi}) \;> 0\;\;.\qquad
\end{equation}
In Subsect.~ \ref{subsect:morphon} we have introduced a mean field description of kinematical backreaction in terms of a
{\em morphon field}. For such an effective scalar field the strong energy condition is not violated for the true content of the
Universe, that is ordinary dust matter. In this line
it is interesting that we can identify `violation' of an effective `strong energy condition' with the
acceleration condition above ({\it cf.} Eqs.~(\ref{equationofstate}), (\ref{morphon:sources})): 
\begin{equation}
\label{energyconditionQ}
3\frac{{\ddot a}_\CD}{a_\CD} =  - 4\pi G (\varrho^{\CD}_{\rm eff}
+3{p}^{\CD}_{\rm eff}) =- 4\pi G\left(\, \langle\varrho\rangle_\CD + 
\varrho_{\Phi}^\CD + 3 p_{\Phi}^\CD\,\right)
= -4\pi G \langle\varrho\rangle_\CD + {\CQ}_\CD \,,
\end{equation}
which has to be positive, if the acceleration condition (\ref{accelerationcondition}) is met.

\subsubsection{Curvature--fluctuation coupling}

It is clear by now that a {\em backreaction--driven cosmology} \cite{rasanen} must make efficient use
of the genuinely relativistic effect that couples averaged extrinsic and intrinsic curvature invariants,
as is furnished by the integrability condition (\ref{integrability}) (or the
Klein--Gordon equation (\ref{kleingordon}) in the mean field description). 
While models that suppress the scalar field degrees
of freedom attributed to backreaction (or the {\em morphon field} in the mean field description),  and so cannot lead to an explanation of Dark Energy on the Hubble scale, general relativity offers a wider range
of possible cosmologies, since it is not constrained by the assumption of Euclidean or
constant--curvature geometry and small deviations thereof. Here, it is essentially the requirement that the evolution of the background geometry
is suppressed (naturally in Newtonian models and through `gauge--fixing' in gauge--invariant perturbation theory), while generically the geometry
is a dynamical variable and {\em does not} evolve independently of the perturbations. 
But, how can a cosmological model be driven away from a `near--Friedmannian' state, if
we do not already start with initial data away from a perturbed Friedmannian
model? How does the {\em mechanism} of the coupling between geometry and matter fluctuations work,
and can this mechanism be sufficiently effective?

\subsubsection{The `Newtonian anchor'}

Let us guide our thoughts by the following intuitive picture.
Integral properties of Newtonian and quasi--Newtonian models
remain unchanged irrespective of whether fluctuations are absent or `turned on'. 
Imagine a ship in a silent water and wind
environment (homogeneous equilibrium state). Newtonian and quasi--Newtonian models do not allow,
by construction \cite{wald}, that the ship would move away as soon as  water and wind become more violent.  
This `Newtonian anchor' is lifted into the ship as soon as we allow for the coupling
of fluctuations to the geometry of spatial hypersurfaces in the form of the averaged
scalar curvature. It is this coupling that can potentially drive the ship away, i.e. change the
integral properties of the cosmology.
Before we are going to exemplify this coupling mechanism, e.g. by discussing  exact solutions, let us add some understanding to the role played by the averaged scalar curvature.

\subsubsection{The role of curvature}

Looking at the integral of the {\em curvature--fluctuation--coupling}, Eq.~(\ref{integrabilityintegral}), we understand that the constant--curvature of the standard
model is specified by the integration constant $k_{\initial\CD}$. This term does not 
play a crucial dynamical role as soon as backreaction is at work. Envisaging a cosmology that
is driven by backreaction, we may as well dismiss this constant altogether. In such a case, the averaged
curvature is {\em dynamically} ruled by the backreaction term and its history.
Given this remark  we must expect that the averaged scalar curvature may experience
changes in the course of evolution (in terms of deviations from constant--curvature), as soon as the structure formation process injects backreaction. 
This picture is actually what one needs in order 
to solve the {\em coincidence problem}, i.e. the observation that the onset of acceleration of the 
Universe seems to coincide with the epoch of structure formation. 

This mechanism can be qualitatively understood by studying {\em scaling solutions}, {\it cf.} Subsect.~\ref{subsubsect:scaling}, which
impose a direct coupling, $\CQ_\CD \propto \average{\CR}$. (These scaling solutions correspond to {\em quiessence fields}, Eq.~(\ref{quiessence}), and have been thoroughly studied by many people working on quintessence (see \cite{sahnistarobinskii}, 
\cite{copeland:darkenergy} and references therein.)   
In the language of a {\em morphon field}, the mechanism perturbs the `virial equilibrium',
Eq.~(\ref{virial}), such that the potential energy stored in the averaged curvature is
released and injected into an excess of kinetic energy (kinematical backreaction).
Thus, in this picture, positive backreaction, capable of mimicking Dark Energy, is fed
by the global `curvature energy reservoir'.
It is clear that such a mechanism relies on an evolution of curvature that differs from the
evolution of the constant--curvature part of the standard cosmology.
Indeed, as we shall exemplify below, already a deviation term of the form  $W_\CD = \average{\CR} - 6 k_{\initial\CD}a_\CD^{-2} \propto a_\CD^{-3}$ is sufficient to change the decay rate of $\CQ_\CD$ from $\propto a_ \CD^{-6}$ to $\propto a_ \CD^{-3}$.

If we start with `near--Friedmannian initial data', and no cosmological constant, then the averaged curvature must be
{\em negative} today and -- if we require the model to fully account for $\Lambda$ -- of the order of the value that we would find for a 
void--dominated Universe \cite{morphon}. 
Thus, the determination of curvature evolution, even only asymptotically \cite{reiris1}, \cite{reiris2}, is key to understand backreaction.
The difference to the concordance model is essentially that the averaged curvature
changes from an almost negligible value at the CMB epoch to a cosmologically relevant 
negative curvature today. This is one of the direct hints to put backreaction onto the stage
of observational cosmology, {\it cf.} Subsect.~\ref{subsubsect:mockmetric}.

Let us add three remarks. First, it is not at all evident that a flat Universe 
is necessarily favoured by the data {\em throughout the evolution} \cite{ichikawa}. This latter
analysis has been performed within the framework of the standard model, and it is clear
that in the wider framework discussed here, the problem of interpreting astronomical data
is more involved. Second, it is often said that spatial curvature can only be relevant near Black Holes
and can therefore not be substantial. Here, one mistakenly implies an astrophysical Black Hole,
while the Schwarzschild radius corresponding to the matter content in a Hubble volume is of
the order of the Hubble scale. As the averaged Hamiltonian constraint (\ref{averagedhamilton}) shows, 
the averaged scalar curvature is a quantitatively competitive player that could only be `compensated' 
(and only on a specified scale) by introducing a cosmological constant. 
In essence, a cosmologically relevant curvature contribution is
tiny, but this property is shared by all cosmological sources.
Third, even standard perturbation theory predicts a scaling--law
for the averaged scalar curvature
that substantially differs from the evolution of a constant--curvature model,
see Subsect.~\ref{subsect:approximations}.

\noindent
(The above qualitative picture is illustrated in detail in R\"as\"anen's review \cite{rasanen}.)

\subsection{Exact solutions for kinematical backreaction}
\label{subsect:exactsolutions}

The following families of exact solutions of the averaged equations are used to illustrate
the mechanism of a backreaction--driven cosmology. Other implications of these examples
are discussed in \cite{buchert:static}.

\subsubsection{A word on the cosmological principle}

We may separate the following classes of solutions into those solutions that respect the cosmological principle and those that do not.
It is therefore worth recalling the assumptions behind the cosmological principle.
In the literature one often finds a `strong' version that
demands {\em local isotropy} of the universe model. More realistically, however,
we should define a `weak' version that refers to  
the existence of a {\em scale of homogeneity}: we assume
that there exists a scale beyond which all observables do no longer depend on scale. It is beyond
this scale where the standard model is supposed to describe the Universe on average; it is
simply unreasonable to apply this model, even on average, to smaller scales, since the 
standard, spatially flat FLRW model has an in--built scale--independence. On the same grounds, isotropy 
can only be expected on the homogeneity scale and not below.
Accepting the existence of this scale has strong implications, one of them being that
cosmological parameters on that scale are {\em representative} for the whole Universe.
If this were not so, and generically we may think of, e.g. a  decay of average
characteristics with scale all the way to the diameter of the Universe as in a generic fractal (or multi--fractal) distribution \cite{joyceetal}, then the cosmological parameters
of the standard model would make no sense unless the scale is explicitly indicated. The homogeneity scale is thought to be well below the 
scale of the observable Universe and within our past--lightcone. Therefore, with this assumption,
averaging over non--causally connected regions
delivers the same values as those already accumulated up to the homogeneity scale \cite{rasanen}, \cite{buchertcarfora:Q}.

We are now briefly describing some exact solutions, and we mainly have in mind to learn about
the coupling between curvature and fluctuations.

\subsubsection{Backreaction as a constant curvature or a cosmological constant}

Kinematical backreaction terms can model a constant--curvature term as is already evident
from the integrability condition (\ref{integrability}). Also, a cosmological constant need not
be included into the cosmological equations, since 
${\CQ}_\CD$ can play this role \cite{buchert:jgrg}, \cite{bks}, \cite{rasanen:constraints},
and can even provide a {\em constant} exactly, as was shown in \cite{kolbetal} and \cite{buchert:static}. The exact condition can 
be inferred from Eq.~(\ref{averagedraychaudhuri}) and (\ref{generalexpansionlaw}) and reads:
\begin{equation}
\label{lambdacondition}
 \frac{2}{ a_\CD^2 }
\int_{\initial{t}}^t \rmd t'\ {\CQ}_\CD\frac{\rmd }{\rmd t'} a^2_\CD(t')
\;\equiv\;{\CQ}_\CD \;\;,
\end{equation}
which implies ${\CQ}_\CD = {\CQ}_{\cal D}(t_i) = const.$ as the only possible 
solution. Such a `cosmological constant' installs, however, via
Eq.~(\ref{integrabilityintegral}), a non--vanishing averaged scalar curvature (even for 
$k_{\initial\CD} =0$):
\begin{equation}
\label{lambdacurvature}
\average{\CR} \;=\;\frac{6 k_{\initial\CD}}{a_\CD^2} \;-\;3 {\CQ}_\CD (t_i)\;\;.
\end{equation}
This fact has interesting consequences for `morphed' inflationary models \cite{morphedinflation}.

\subsubsection{The Universe in an out--of--equilibrium state: a fluctuating Einstein cosmos}

Following Einstein's thought to construct a globally static model, we 
may require the effective scale--factor $a_{\Sigma}$ on a simply--connected 3--manifold $\Sigma$ without boundary to be 
constant on some time--interval, hence ${\dot a}_{\Sigma} = {\ddot a}_{\Sigma}=0$ and
Eqs.~(\ref{averagedraychaudhuri}) and (\ref{averagedhamilton}) 
may be written in the form:
\begin{equation}
\label{static}
{\CQ}_{\Sigma}\;=\;4\pi G \frac{M_{\Sigma}}{\initial{V}a_{\Sigma}^3} - \Lambda\;\;\;;\;\;\;
\gaverage{\CR}\;=\; 12\pi G \frac{M_{\Sigma}}{\initial{V}a_{\Sigma}^3}+3\Lambda\;\;,
\end{equation}
with the global {\it kinematical backreaction} ${\CQ}_{\Sigma}$, the globally averaged scalar 3--Ricci 
curvature $\gaverage{\CR}$, and the total restmass  $M_{\Sigma}$ contained in
$\Sigma$.

Let us now consider the case of a vanishing cosmological constant: $\Lambda = 0$.
The averaged scalar curvature  is, for a non--empty Universe, 
always {\it positive}, and 
the balance conditions (\ref{static}) replace Einstein's balance 
conditions that determined the cosmological constant in the standard 
homogeneous Einstein cosmos. 
A globally static inhomogeneous cosmos without a cosmological constant 
is conceivable and characterized by the cosmic equation of state:
\begin{equation}
\label{cosmicstate1}
\gaverage{\CR}\;=\;3{\CQ}_{\Sigma}\;=\;const.\;\;\Rightarrow\;\;
{p}^{\Sigma}_{\rm eff}\;=\; \varrho^{\Sigma}_{\rm eff}\;=\;0\;\;.
\end{equation}
Eq.~(\ref{cosmicstate1}) is a simple example of a strong coupling between curvature and
fluctuations. Note that, in this cosmos, the effective Schwarzschild radius
is larger than the radius of the Universe,
\begin{equation}
\label{blackhole}
a_{\Sigma} = \frac{1}{\sqrt{4\pi G \gaverage{\varrho}}}\;=\;\frac{1}{\pi} 2 G M_{\Sigma}= \frac{1}{\pi} a_{\rm Schwarzschild}\;,
\end{equation}
hence confirms the cosmological relevance of curvature on the global scale $\Sigma$.
The term `out--of--equilibrium' refers to 
our measure of relative information entropy, {\it cf.} Subsect.~\ref{subsect:information}: in the above
example volume expansion cannot compete with information production because the volume
is static, while information is produced (see \cite{buchert:static} for more details).

Such examples of global restrictions imposed on the averaged equations do not refer to 
a specific inhomogeneous metric, but should be thought of in the spirit of the virial theorem
that also specifies integral properties but without a guarantee for the existence of
inhomogeneous solutions that would satisfy this condition. (In \cite{buchert:static} a possible stabilization mechanism of a stationarity condition
by backreaction, as
opposed to the global instability of the classical Einstein cosmos,
has been discussed.)

\subsubsection{Demonstration of the backreaction mechanism: \\\quad\quad\; a globally stationary inhomogeneous cosmos}

Suppose that the Universe indeed is hovering around 
a non--accelerating state on the largest scales. 
A wider class of models that balances the fluctuations and the 
averaged sources can be constructed by introducing 
{\it globally stationary effective cosmologies}: the vanishing of the second time--derivative
of the scale--factor would only imply ${\dot a}_{\Sigma} = const.=:{C}$, i.e.,
$a_{\Sigma} = a_S + {C}(t-t_i)$, where the integration constant $a_S$
is generically non--zero, e.g. the model may emerge \cite{ellis:emergent1}, \cite{ellis:emergent2}
from a globally static cosmos, $a_S :=1$, or from a  
`Big--Bang', if $a_S$ is set to zero. 
In this respect this cosmos does not appear very different from the standard model, since it evolves at an effective Hubble rate $H_{\Sigma} \propto 1/t$. (There are, however, substantial
differences in the evolution of cosmological parameters, see \cite{buchert:static}, Appendix B.)

The averaged equations
deliver a dynamical coupling relation between  
${\CQ}_{\Sigma}$ and $\gaverage{\CR}$ as a special case of the integrability condition (\ref{integrability})\footnote{The constant $C$ is determined, for the normalization $a_{\Sigma}(t_i) =1$, by: \\$6{C}^2 = 6\Lambda + 3{\CQ}_{\Sigma}(t_i) - \gaverage{\CR}(t_i)$.}:
\begin{equation}
\label{coupling}
-\partial_t {\CQ}_{\Sigma} + \frac{1}{3} \partial_t  \gaverage{\CR}\;=\;
\frac{4{C}^3}{a_{\Sigma}^3}\;.
\end{equation}
The cosmic equation of state of the $\Lambda-$free stationary cosmos and its solutions read \cite{buchert:darkenergy,buchert:static}:
\begin{eqnarray}
\label{cosmicstate3}
{p}^{\Sigma}_{\rm eff}\;=\; 
-\frac{1}{3}\;\varrho^{\Sigma}_{\rm eff}\;\;\;\;;\;\;\;\;
\label{stationaryS}
{\CQ}_{\Sigma} \;=\; \frac{{\CQ}_{\Sigma}(t_i)}{a_{\Sigma}^{3}}\;;\\
\label{curvatureevolutionR}
\gaverage{\CR}=  \frac{3 {\CQ}_{\Sigma}(t_i)}{a_{\Sigma}^3} - 
\frac{3{\CQ}_{\Sigma}(t_i)-\gaverage{\CR}(t_i)}{a_{\Sigma}^2}\;.
\end{eqnarray}
The total kinematical backreaction ${Q}_{\Sigma}V_{\Sigma} = 4\pi G M_{\Sigma}$ 
is a conserved quantity in this case.

The stationary state tends to the
static state only in the sense that, e.g. in the case of an expanding cosmos, 
the rate of expansion slows down, but the steady increase of the scale factor allows for a global
change of the sign of the averaged scalar curvature.   
As Eq.~(\ref{curvatureevolutionR}) shows, an initially
positive averaged scalar curvature would decrease, and eventually would become 
negative as a result of backreaction.
This may not necessarily be regarded as a signature of  a global topology change, as 
a corresponding sign change in a Friedmannian model would suggest (see
Subsect~\ref{subsect:global}). 

The above two examples of globally non--accelerating universe models evidently violate the cosmological principle, while they would
imply a straightforward explanation of Dark Energy on regional (Hubble) scales:
in the latter example the
averaged scalar curvature has acquired a piece $\propto a_{\Sigma}^{-3}$ that, astonishingly,
had a large impact on the backreaction parameter, changing its decay rate from
$\propto a_{\Sigma}^{-6}$ to $\propto a_{\Sigma}^{-3}$, i.e. the same decay rate as that of the averaged density. This is certainly enough to 
produce sufficient `Dark Energy' on some regional patch due to the presence of strong 
fluctuations\footnote{In \cite{buchert:static} a conservative estimate, based on currently discussed numbers for the cosmological parameters,
shows that such a cosmos provides room for at least $50$ Hubble volumes.} \cite{buchert:darkenergy}. However, solutions that respect the cosmological principle and, at the
same time, satisfy observational constraints can also be
constructed \cite{morphon}. In this latter work, scaling solutions that we shall discuss now, have been exploited for such 
a more conservative approach.

\subsubsection{The solution space explored by scaling solutions}
\label{subsubsect:scaling}

In \cite{morphon} a systematic classification of scaling solutions of the averaged equations
was given. Like the averaged dust matter density 
$\langle\varrho\rangle_\CD$ that evolves, for a restmass preserving domain $\CD$, as 
$\langle\varrho\rangle_\CD=\langle\varrho\rangle_{\initial\CD}\; a_\CD^{-3}$, we can look at the case
where also  the backreaction term and the averaged scalar curvature obey scaling laws,
\begin{equation}
\label{prescription}
{\CQ}_\CD={\CQ}_{\initial\CD}\; a_\CD^{n}\;\;\;\;;\;\;\;\;
\average{\CR}={\CR}_{\initial \CD}\; a_{\CD}^{p}\;\;,
\end{equation}
where ${\CQ}_{\initial\CD}$ and ${\CR}_{\initial\CD}$ 
denote the initial values of ${\CQ}_\CD$ and $\average{\CR}$, respectively.
The integrability condition (\ref{integrability}) then immediately provides as 
a first scaling solution (\cite{buchert:grgdust}, Appendix B):
\begin{equation}
\label{ndiffpsolution}
{\CQ}_\CD={\CQ}_{\initial\CD }\; a_\CD^{-6}\;\;\;\;;\;\;\;\;
\average{\CR}={\CR}_{\initial\CD}\; a_\CD^{-2}\;\;.
\end{equation} 
This is the only solution with $n\neq p$.
In the case $n=p$, we can define a coupling parameter $r_\CD$ (that can be chosen
differently for a chosen domain of averaging\footnote{For notational ease we henceforth
drop the index $\CD$ and simply write $r$.}) such that ${\CQ}_{\initial\CD}\propto
{\CR}_{\initial\CD}$; the solution reads: 
\begin{equation}
\label{backreactionparameter}
{\CQ}_\CD \;=\;r \;\average{\CR}\;=\;r \;{\CR}_{\initial\CD} \,a_\CD^{n}\;\;\;\;\;;\;\;n=-2 \frac{(1+3r)}{(1+r)}\;\;;\;\;r= - \frac{(n+2)}{(n+6)}\;\;,
\end{equation}
(with $r\neq -1$ and $n\neq -6$).
The mean field description of backreaction, Subsect.~\ref{subsect:morphon}, defines a scalar field evolving in a {\it positive} potential, if ${\CR}_{\initial\CD}<0$ (and in a {\it negative} potential if ${\CR}_{\initial\CD}>0$), 
and a {\it real} scalar field, if $\epsilon{\CR}_{\initial\CD}(r+1/3)<0$. 
In other words, if ${\CR}_{\initial\CD}<0$ we have a priori a phantom field
for $r<-1/3$ and a standard scalar field for $r>-1/3$; if  ${\CR}_{\initial\CD}>0$, we have a standard scalar field for $r<-1/3$ and a
phantom field for  $r>-1/3$.

For the scaling solutions the explicit form of the self--interaction term of the scalar field can be reconstructed \cite{morphon}: 
\begin{equation}
\label{pot1}
U(\Phi_\CD , \laverage{\varrho})= \frac{2(1+r)}{3}\left((1+r) \frac{\Omega_{\CR}^{\initial\CD}}{\Omega_m^{\initial\CD}} \right)^{\frac{3}{n+3}} 
\langle\varrho\rangle_{\initial\CD}\,
\sinh^{^{\frac{2n}{n+3}}}\left(\frac{(n+3)}{\sqrt{-\epsilon n}}\sqrt{2\pi G} \Phi_\CD \right),
\end{equation}
where $\langle\varrho\rangle_{\initial\CD}$ is the initial averaged restmass density of 
dust matter, introducing a natural scale into the scalar field dynamics.
This potential is well--known in the context of phenomenological quintessence models,
\cite{sahnietal}, \cite{alametal}, \cite{liddle98}, \cite{Peebles2003} and references therein.
The scaling solutions correspond to specific scalar field models with a constant 
fraction of kinetic and potential energies of the scalar field, i.e. with Eq.~(\ref{quiessence}),
\begin{equation}
\label{virial}
E^\CD_{\rm kin}+\frac{(1+3r)}{2\epsilon}E^\CD_{\rm pot}\;=\;0\;\;.
\end{equation}
We previously discussed the case $r=0$ (`zero backreaction') for which this 
condition agrees with the standard scalar virial theorem.

We turn now to an explicit discussion of these scaling solutions summarized in a {\em cosmic phase diagram} in Figure~\ref{fig:cosmicstates}.

\begin{figure}[htbp]
\begin{center}
\includegraphics[width=11.7cm,height=11.7cm]{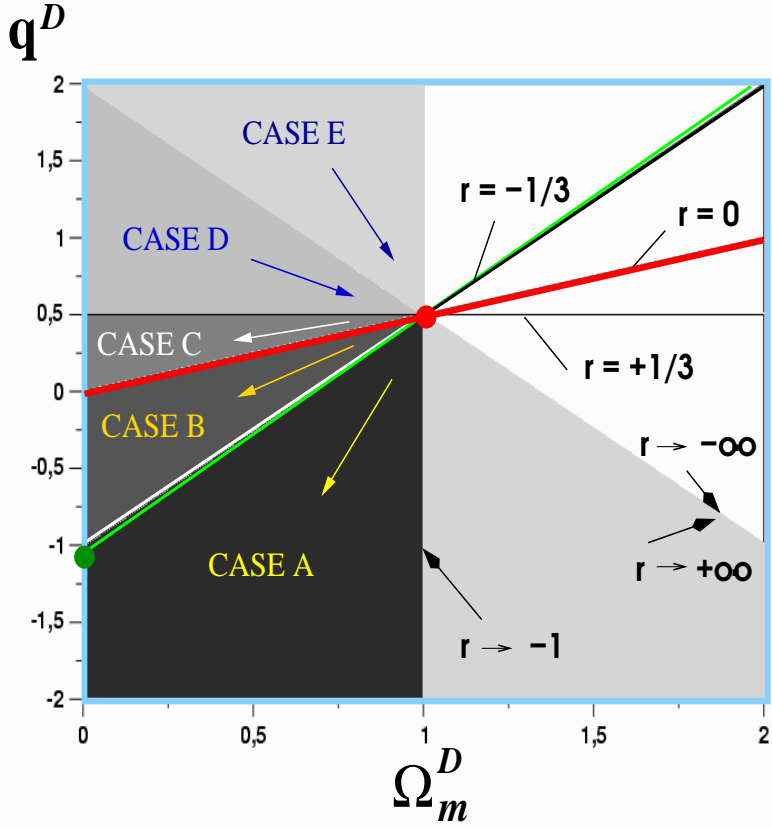}
\caption{\label{fig:cosmicstates}
This `cosmic phase diagram', spanned by the effective volume deceleration parameter $q^\CD$, Eq.~(\ref{deceleration}), and the effective density parameter $\Omega^\CD_m$, Eq.~(\ref{omega}),
is valid for all times and on all scales, i.e. it can be read as 
a diagram for the corresponding parameters `today' on the scale of the observable 
Universe.  It represents a two--dimensional subspace $\lbrace\;\Lambda =0\;\rbrace$  of the full solution space that would include a cosmological constant.
All the scaling solutions are represented by straight lines passing
through the Einstein--de Sitter model in the center of the diagram ($1/2;1$). 
The vertical line corresponding to ($q^{\CD};1$) is not associated with a solution of the 
backreaction problem; it degenerates to the Einstein--de Sitter model ($1/2
;1$). This line forms a `mirror': inside the cone (Case E) there  are
solutions with $\Omega^{\CD}_m>1$ that cannot be related to any real--valued scalar field,
but are still of physical interest in the backreaction context (models with positive averaged
scalar curvature).
Models with `Friedmannian kinematics', but with renormalized parameters form the line 
$r=1/3$ (for details see \cite{morphon}, Appendix A). The line $r=0$ are models with no backreaction 
on which the parameter $\Omega^\CD_k$ varies
(scale--dependent `Friedmannian models').
Below the line $r=0$ in the `quintessence phase'
we find effective models with subdominant shear fluctuations
(${\CQ}_\CD$ positive, $\Omega^\CD_{\CQ}$ negative).The line $r=-1/3$
mimics a `Friedmannian model' with scale--dependent cosmological constant.
The line below $r=-1/3$ in the `phantom quintessence phase'
represents the solution inferred from SNLS data ({\it cf.} \cite{morphon}),
and the point at ($q^{\CD};\Omega^\CD_m) = (-1.03; 0)$ 
locates the late--time attractor associated with this solution. Since we have no cosmological constant here, all expanding solutions in the subplane $q^\CD <0$  drive the averaged variables away from the standard model featuring a backreaction--driven volume acceleration of effectively isotropic cosmologies that are curvature--dominated at late times.}
\end{center}
\end{figure}

\subsubsection{Discussion of Figure~\ref{fig:cosmicstates}}

In Figure~1 we only concentrate on the two--dimensional solution space of averaged inhomogeneous cosmologies without
a cosmological constant. We further concentrate in this discussion only on expanding universe models; the solution space contains also contracting
models that are equally relevant if we interpret this figure for smaller spatial scales; (recall that we have ${\CR}_{\initial\CD}<0$ for $r>-1$ and ${\CR}_{\initial\CD}>0$ for $r<-1$).

A phase space analysis of the scaling solutions \cite{morphon} shows that the Einstein--de Sitter model is a
saddle point for the scaling dynamics and small inhomogeneities with 
${\CQ}_{\CD}>0$ should make the system evolve away from it.
The sign of ${\CQ}_{\CD}$ is important: for all the
models corresponding to $r>0$ or $r<-1$, that is the cases C,D and E in Figure~\ref{fig:cosmicstates}, which cannot
produce accelerated expansion, we have ${\CQ}_{\CD}<0$. In other words, the
kinematical backreaction is dominated by shear fluctuations, {\it cf.} 
Eq.~(\ref{backreactionterm}). This does not necessarily mean that the
universe model is regionally (on the scale $\CD$) anisotropic, because in these 
cases kinematical fluctuations decay rapidly.  On the other hand,
cases A and B that could be responsible for an accelerated expansion
correspond to ${\CQ}_{\CD}>0$ and have subdominant shear fluctuations. 
Therefore, these models can be regionally almost 
isotropic, although kinematical fluctuations have strong influence.

Moving down the cases from Case E to Case A we first have
models in which ${\CQ}_{\CD}$ decays stronger than the density; equal decay rate ${\CQ}_{\CD}\propto a_\CD^{-3}$ is found on the line $r=1/3$.
This situation changes for Case C where the Friedmannian kinematics does no longer act as an attractor: backreaction,
having a decay rate weaker than the density, entails an averaged curvature evolution that deviates from a constant--curvature Friedmannian model.
Case B represents the quintessence phase in the scalar field correspondence, in which the averaged model accelerates, bounded below by the line $r=-1/3$ of a constant backreaction (exactly modeling a cosmological constant on a given scale). While fitting supernovae data with a constant negative curvature
(the line $r=0$ left to the Einstein--de Sitter model) is not successful, we nevertheless appreciate that such Friedmannian models would physically
mimic the instability towards a curvature--dominated phase. Deviations from constant--curvature carry the averaged model into the quintessence or
even phantom quintessence regime (Case A), in which case backreaction is growing (as seen within the on average negatively curved space!).
In Section~4, Subsect.~\ref{subsubsect:perturbations}, we shall discuss a perturbative model that features as a leading mode a decay rate ${\CQ}_{\CD}\propto a_\CD^{-1}$ with a
deviation from constant--curvature at the same rate, $\average{\CR} \propto a_\CD^{-1}$. This (conservative) model already lies in the
quintessence phase of an accelerating universe model and can be located on the line $r= -1/5$ in between the constant--curvature line and the
`cosmological constant'. Thus, in this figure and explicitly in Figure~\ref{fig:zoom}, an explanation of Dark Energy through backreaction effects is expressed by the expectation that
a non--perturbative model would weaken the leading perturbative mode further; it would certainly lie below ${\CQ}_{\CD}\propto a_\CD^{-1}$. 
We shall continue this discussion in the context of perturbative solutions in Subsect.~\ref{subsubsect:perturbations}.

\subsubsection{Explicit inhomogeneous solutions}

If we wish to specify the evolution of averaged quantities without resorting to phenomenological assumptions on the equations of state of the various ingredients, or on
the necessarily qualitative analysis of scaling solutions, or with specific global assumptions,
we have to specify the inhomogeneous metric \cite{krasinski}.
Natural first candidates are the spherically--symmetric
Lema\^\i tre--Tolman--Bondi (LTB) solutions that were first employed in the context
of backreaction in \cite{celerier1} and \cite{rasanen:LTB}.

Considerable effort has been spent on LTB solutions and, especially recently, relations to 
integral properties of averaged cosmologies have been sought. Interestingly, 
\cite{nambu} also found a strong coupling between averaged scalar curvature
and kinematical backreaction, and LTB solutions also feature an additional curvature piece  $\propto a_{\CD}^{-3}$ on some domain $\CD$. There are obvious shortcomings of LTB solution studies that consider the class of on average vanishing scalar curvature, since in that
class also $\CQ_\CD \equiv 0$ \cite{singh1};
also here, a non--vanishing averaged curvature is crucial to study backreaction \cite{chuang}.
However, there is enough motivation to quantify the extra effect of a positive expansion variance
to fit observational data (\cite{celerier2} and references below).

The value of LTB studies or studies of other highly symmetric exact solutions is more to be seen in the specification of 
observational properties such as the luminosity distance in an
inhomogeneous metric \cite{LTBgron,LTBluminosity5,LTBluminosity1,LTBluminosity2,LTBluminosity4,alnes2,alnes3}, as well as Enqvist (this volume).
Although interesting results were obtained, especially in connection with the interpretation
of supernova data, care must be taken when determining e.g. just luminosity distances, since
the free LTB functions may fit any data \cite{mustapha}. 
Generally, apart from mistakes (e.g. setting the shear to zero), those studies sometimes
confuse integral properties of a cosmological model with local properties (e.g. the scale factor $a_\CD$ and a local scale factor in the given metric form). The averaged equations cannot
predict luminosity distances unless one considers averages on the lightcone, {\it cf.} Subsect.~\ref{lightcone} (see, however,
different strategies proposed and pursued in \cite{palle}), \cite{bolejko}, and \cite{teppo2}),  which in turn
is related to the issue of light--propagation in an inhomogeneous Universe
(see \cite{kantowski}, \cite{kantowski2}, \cite{kantowski3}, \cite{sugiura}, \cite{ruth:luminosity}, \cite{kolbetal_swisscheese}, and discussion and 
references in \cite{ellisbuchert}).
A promising strategy to exploit the LTB solution is to consider an ensemble of spherical regions whose initial data are constrained
by a standard Cold Dark Matter power spectrum, and to look at the correlated average properties of the ensemble\footnote{R\"as\"anen ({\it priv. comm.})
is currently looking at an ensemble of spherical regions in the spherical collapse model to describe the statistical distribution of expanding and collapsing regions, where the statistical properties of this ensemble are given by the peak model of structure formation for CDM.}. However, in order to avoid 
matching conditions that are necessarily involved for an ensemble of LTB solutions, a generic collapse model in the spirit of the Newtonian model
investigated in \cite{abundance} would facilitate such a description.

Another possibility to construct explicit inhomogeneous metrics is, of course, to employ perturbative, but also non--perturbative assumptions, that will be both discussed in Subsection~\ref{subsect:approximations}.

\newpage

\section{Future theoretical and observational strategies:\\ $\quad$--- quantitative views on backreaction}
\label{subsect:outlook}
In this section we are going to outline several strategies towards the goal of 
understanding the quantitative importance of backreaction effects, and to device methods of their
observational interpretation. All the topics discussed below are the subject of  work in progress.

\subsection{Global aspects}
\label{subsect:global}

The question of what actually determines the averaged scalar curvature is open.
For a two--dimensional Riemannian manifold this question is answered through the
Gauss--Bonnet theorem: the averaged scalar curvature is determined by the Euler--characteristic
of the manifold. Hence, {\em it is a global topological property rather than a certain restriction
on local properties of fluctuations that determines the averaged scalar curvature}. 
If such an argument would carry over to a three--dimensional manifold, then any local argument for an estimate of backreaction would obviously be off the table. (There are related thoughts and results in
string theory that could be very helpful here.)
In ongoing work \cite{carforabuchert:perelman,buchertcarfora:Q} we consider the consequences of Perelman's work that was mentioned 
in Subsect.~\ref{subsubsect:ricciflow}. 
There is no such theorem like that of Gauss and Bonnet
in three dimensions, but there are uniformization theorems that could provide similar conclusions.
For example, for closed inhomogeneous universe models we can apply Poincar\'e's conjecture (now proven by Perelman \cite{perelman:entropy}, \cite{perelman:ricci}) that any simply--connected three--dimensional Riemannian manifold without boundary  is a homeomorph of a 3--sphere. 
Ongoing work concentrates on the multi--scale analysis of the curvature distribution and the related distribution of
kinematical backreaction on cosmological hypersurfaces that feature the phenomenology we
observe. All these studies underline the relevance of topological issues for a full understanding
of backreaction in relativistic cosmology. To keep up with the developments in 
Riemannian geometry and related mathematical fields will be key to advance 
cosmological research. 
In this line it should be stressed that the averaged scalar curvature is only a weak descriptor for the topology in the general 3D case, and information on the sectional curvatures or the full Ricci tensor is required.
In observational cosmology there are already a number of efforts,
e.g. related to the observation of the topological structure of the Universe derived from CMB maps (for further discussion see \cite{buchert:static} and for topology--related issues see 
\cite{uzan:topology,lehoucq:topology}, \cite{cornishspergel}, \cite{aurich1:topology,aurich2:topology}, \cite{tavakol:topology}). 

\subsection{Perturbative and non--perturbative approaches to backreaction}
\label{subsect:approximations}

There is a large body of possibilities to construct a generic inhomogeneous metric.
First, there is the possibility of using standard methods of perturbation theory.
Although the equations and `parameters' discussed in this work can live without introducing
a background spacetime, a concrete model for the backreaction terms can be obtained by
employing perturbation theory (preferably of the Lagrangian type) and, hence, a reference background must be introduced. But, the construction idea is (i) {\em to only model
the fluctuations by perturbation theory (the term $\CQ_\CD$) and to find the final (non--perturbative) model 
by employing the exact framework of the averaged equations}. Such a model is currently 
investigated by paraphrasing the corresponding Newtonian approximation \cite{bks}. 
We shall outline more in detail below what we expect to learn from such a model.
Second, we could aim at finding an approximate evolution equation for $\CQ_\CD$ 
by (ii) {\em closing the hierarchy of ordinary differential equations} that involve the evolution of shear
and the electric and magnetic parts of the projected Weyl tensor. 
The problem of closing such a hierarchy of equations is often considered in the 
literature and various closure conditions are formulated 
(e.g., \cite{bertschinger:hui}). One of them, the {\em silent universe model} \cite{bruni:silent},
which assumes a vanishing magnetic part of the Weyl tensor, 
is found to be too restrictive to describe a realistic inhomogeneous Universe \cite{henk:silent}, \cite{silent2}, \cite{silent3},
so that we need to head for closures with non--vanishing magnetic part.
In this line, (iii) {\em further studies of cosmic equations of state} (like, e.g., the Chaplygin 
state \cite{chaplygin})  are not only a clearcut way to close the averaged equations, but also a way to classify different solution sectors. 
All these models could be subjected to (iv) {\em standard dynamical system's analysis to show their
stability in the phase space of their parameters} \cite{wainwrightellis,henk:attractor}. 

As already remarked above, the FLRW cosmology as an averaged model is found to be stable in many cases,
but there is an unstable sector that just lies in the right corner needed to explain `Dark Energy'.
In order to analyze this instability, we first look at perturbation theory in Lagrangian form.
The following excursion allows us to roughly examine the possibilities provided by perturbation theory and to
identify the unstable mode that is of interest in the Dark Energy context, although we do not expect such an approach
to be sufficient. We shall also begin to investigate non--perturbative methods below.

\subsubsection{Relativistic Lagrangian perturbation theory}
\label{subsubsect:RZA}

The following is a shortcut to a setup that will provide insights without entering a detailed perturbative analysis.
The idea is to generalize the Newtonian results on backreaction, investigated in detail in \cite{bks}.
For this purpose it is enough to note that in a comoving and synchronous setting the electric part of the projected Weyl tensor 
is sufficient to capture the relativistic generalization of a first--order Lagrangian perturbation scheme in Newtonian cosmology. 
This latter is furnished by a Lagrangian set of evolution equations for a family of trajectories, sending an initial (Lagrangian) position vector $X^i$ to its
Eulerian  position vector at time $t$, $x^i = {\bf f} (X^i , t)$ in a Euclidean embedding space.
The relativistic generalization of the exact spatial one--forms $dx^i$ is provided by Cartan co--frame fields 
$\boldsymbol{\eta}^a = \eta^a_{\;\,j}dX^j$\footnote{The indices ($a,b,c ...$) are here non--coordinate indices that just count the one--forms,
as opposed to the coordinate indices ($i,j,k ...$).} deforming
the local exact basis $dX^j$. Correspondingly, the first--order Lagrangian perturbation solution \cite{buchert92} $f^i = a(t) X^i + \xi (t) P^i (X^i )$,
with $a(t)$ solving the standard Friedmann equations and $\xi (t)$ a background--dependent known function of time, has its analog
in the relativistic deformation one--form $\boldsymbol{\eta}^a = a(t) {\bf X}^a + \xi (t) {\bf P}^a(X^i )$ \cite{kasai95}, \cite{matarrese&terranova}. 
This approximation solves the 
`electric part' of the projected Einstein equations, written for Cartan co--frame fields, to first order. This part of Einstein's equations, consisting of four
equations for the nine co--frame coefficients $\eta^a_{\;\,i}$ with determinant
\begin{equation}
\label{determinantGR}
J: =\det (\eta^a_{\;\,i}) = \frac{1}{6}
\epsilon_{abc}\epsilon^{ijk}\,\eta^a_{\;\,i}\eta^b_{\;\,j}\eta^c_{\;\,k}\;\;,
\end{equation}
can be written \cite{buchertostermann}, \cite{buchert:cup}: 
\begin{equation}
\delta_{ab} \;{\ddot\eta}^a_{\;[j}\, \eta^b_{\;i]} \;=\; 0\;\;\;\;;\;\;\;\;
\frac{1}{2}\,\epsilon_{abc}\,\epsilon^{ijk}
\; {\ddot\eta}^a_{\;i}\, {\eta}^b_{\;j}\, {\eta}^c_{\;k}\;
=\Lambda J-{4\pi}G\varrho_i (X^i )\;\;.
\label{LESelectricpart}
\end{equation}
This system of equations is the relativistic
(non--Euclidean) generalization of the {\it Lagrange--Newton system} (\ref{LNS}) below 
for {\it dust matter}\footnote{This (closed) system of equations was obtained in \cite{buchertgoetz} for the case of no background source, 
in particular $\Lambda=0$, and in \cite{buchert89} 
including backgrounds of Friedmann--Lema\^\i tre type. The function $\xi (t)$ is given for backgrounds including $\Lambda$ in \cite{bildhaueretal}. A review and alternative forms of these equations may be found in \cite{ehlersbuchert}.}:
\begin{equation}
\delta_{ij} \;{\ddot f}^i_{\;[\vert j}\, f^j_{\;\vert i]} \;=\; 0\;\;\;\;;\;\;\;\;
\frac{1}{2}\,\epsilon_{\ell mn}\,\epsilon^{ijk}
\; {\ddot f}^{\ell}_{\;\vert i}\, {f}^m_{\;\vert j}\, {f}^n_{\;\vert k}\;
=\Lambda J-{4\pi}G\varrho_i (X^i )\;\;.
\label{LNS}
\end{equation}
The geometrical limit that sends the non--exact Cartan forms to the exact forms $d f^i$ (implying that the metric of space is flat) 
reduces the system (\ref{LESelectricpart}) to the Newtonian system (\ref{LNS}), demonstrating that the comoving synchronous spacetime slicing considered has a clearcut Newtonian limit\footnote{A rigorous account for this Newtonian limit, employing the full set of Einstein's equations that
includes the `magnetic part', will be given in \cite{buchertostermann} and \cite{buchert:cup}.
In a post--Newtonian setting the Newtonian limit leads to the Eulerian representation of the Newtonian system, while in the comoving setting considered here it leads to its Lagrangian representation.}.

\subsubsection{A non--perturbative model for backreaction and the leading mode}
\label{subsubsect:perturbations}

Combined with the relativistic form of Zel'dovich's model \cite{zeldovich1}, \cite{zeldovich2}, \cite{buchert92}, straightforward generalization of the results provided in \cite{bks} yields a backreaction term that separates
into its time--evolution given  by $\xi(t)$ and the spatial dependence
on  the  initial  displacement   field  given  by  averages  over  the principal scalar
invariants of the extrinsic curvature coefficients at initial time, $\initial{\inI}, \initial{\inII}, \initial{\inIII}$:
\begin{equation}
\label{eq:Q-full-zel}
{\CQ}_\CD  = \frac{\dot{\xi}^2 \;(\Upsilon_1 + \xi \Upsilon_2 + \xi^2 \Upsilon_3)}{
\left(1 + \xi\laverage{\initial{\inI}} + \xi^2\laverage{\initial{\inII}}
 + \xi^3\laverage{\initial{\inIII}} \right)^{2}}\;\;,
\end{equation}
with $\,\Upsilon_1 := 2\laverage{\initial{\inII}} - \frac{2}{3}\laverage{\initial{\inI}}^2 \,=\,\CQ_{\initial\CD}\,$, and
\begin{equation*} 
\Upsilon_2 :=
6\laverage{\initial{\inIII}}- \frac{2}{3}\laverage{\initial{\inI}}\laverage{\initial{\inII}}
\;\;\;\;;\;\;\;\;
\Upsilon_3 :=2\laverage{\initial{\inI}}\laverage{\initial{\inIII}}
  -\frac{2}{3}\laverage{\initial{\inII}}^2 \;\;.
\end{equation*}
The   first  term in the numerator is  global and  corresponds  to the
linear   damping   factor;   in   an   Einstein--de--Sitter   universe
$\dot{\xi}^2 \propto a^{-1}$.  The denominator  of the first term is a
volume  effect,  whereas the  second  term  in  brackets features  the
initial backreaction as a leading term.

In the  early stages of  structure formation with $\xi(t)\ll1$  we get
\begin{equation}
{\CQ}_\CD  \;\approx\; \frac{1}{a} {\CQ}_\initial{\CD}\;\;,
\end{equation}
identical to the perturbative evolution of $\CQ_\CD$, functionally evaluated with the linear approximation.  
In the Newtonian investigation \cite{bks} it was found that this latter solution is in very good accord
with the general model corresponding to (\ref{eq:Q-full-zel}) on scales larger than $\approx 300$Mpc/h, which entitles us to 
expect that, on large scales, a perturbative model for $\CQ_\CD$ can at best moderately improve on this solution by going to higher orders in the
perturbation scheme. Since $\CQ_\CD$ is quadratic, this mode appears in a relativistic second--order perturbation solution as the leading
mode \cite{rasanen:darkenergy},  \cite{kolbetal}, \cite{lischwarz}, although this leading term is dismissed due to its property to be a complete
divergence in a standard perturbative setting\footnote{Notice that in our derivation of the large--scale behavior of a non--perturbative Lagrangian
model, this is not the case, in agreement with the general situation in a relativistic setting. The backreaction term
is a complete divergence only, if the initial data have this property. This latter is only possible for initially Euclidean geometry.}.
Exploiting the fact that on large scales we only find a small deviation of the volume scale factor 
$a_\CD$ from the Friedmannian scale factor $a(t)$ in this scheme, we may use the exact scaling solution, {\it cf.} Subsect.~\ref{subsubsect:scaling}, 
${\CQ}^S_\CD \propto a_\CD^{-1}$
as a (conservative) prototype model for backreaction, arising as a first leading perturbation in the vicinity of a standard FLRW model.
The averaged scalar curvature corresponding to this scaling solution also evolves with the same power $\average{\CR}^S \propto a_\CD^{-1}$, which
again is in accord with the leading second--order perturbative term found in \cite{lischwarz}. 

\subsubsection{Can backreaction compete with a cosmological constant?}
\label{subsubsect:competitive}

Let us now look at the dimensionless characteristics (\ref{omega}).
For the perturbative scaling modes ${\CQ}^S_\CD$ and $\average{\CR}^S$ discussed in the last subsection we find $\Omega^\CD_{\CQ^S} = -1/5 \Omega^\CD_{\CR^S}$, both are {\em growing} functions of $a_\CD$,
and the relevant term that can play the role of Dark Energy, see Eq.~(\ref{parameterrelation}), divided by the mass density parameter, 
is also growing,
\begin{equation}
\label{om1}
\frac{\Omega^\CD_{\CQ^S} +\Omega^\CD_{\CR^S}}{\Omega^\CD_m}(t) = \frac{-4\Omega^{\initial\CD}_{\CQ}}{\Omega^{\initial\CD}_m}\;   a^2_\CD (t) = \frac{{\CQ}_\initial{\CD}}{4\pi G \laverage{\varrho}}\;   a^2_\CD (t)\;\;\;;
\;\;\;\Omega^\CD_{\Lambda} =0 \;\;,
\end{equation}
clearly demonstrating the (global) instability of the standard model.
This has to be compared with the corresponding fraction of a cosmological constant parameter with respect to the density parameter,
\begin{equation}
\label{om2}
\frac{\Omega^\CD_{\Lambda}}{\Omega^\CD_m} (t) =\frac{\Omega^{\initial\CD}_{\Lambda}}{\Omega^{\initial\CD}_m} \; a^3_\CD (t)
= \frac{\Lambda}{8\pi G \laverage{\varrho}}  \; a^3_\CD (t)\;\;\;;\;\;\;\Omega^\CD_\CQ = 0 \;\;,
\end{equation}
where, with the last assumption, the index of domain--dependence is redundant.
Looking at the respective deceleration parameters, 
\begin{equation}
\label{dec1}
q^\CD_{\CQ^S} =  \frac{1}{2}
\Omega_m^{\CD} + 2 \Omega_{\CQ^S}^{\CD} \;\;\;\;;\;\;\;\;q_{\Lambda}=\frac{1}{2}\Omega_m - \Omega_{\Lambda}\;\;,
\end{equation}
we find in both models the onset of acceleration ($q^\CD_{\CQ^S} = q_{\Lambda} = 0$) at the time when
\begin{equation}
\label{dec2}
a_\CD^{\rm acc} ({\CQ^S}) = \left[\,\frac{4\pi G \laverage{\varrho}}{{\CQ}_{\initial\CD}}\,\right]^{1/2} \;\;\;\;;\;\;\;\;
a^{\rm acc} ({\Lambda}) = \left[\,\frac{4\pi G {\varrho}_H (t_i)}{\Lambda}\,\right]^{1/3}\;\;.
\end{equation}
Although the leading second--order perturbative mode discussed here in the form of a scaling solution lies in the {\em quintessence sector}, {\it cf.} Figures~\ref{fig:cosmicstates} and \ref{fig:zoom}, 
perturbation theory is restricted to a regime close to the Friedmannian state and so, strictly, does not allow
us to follow the scaling mode further towards a curvature--dominated regime. 
However, by extrapolating the scaling behavior of the perturbative mode into this regime, its impact is in principle competitive, even if we set out standard initial data for ${\CQ}_\initial{\CD}$: the comparison of scaling behaviors
of (i) the averaged density, being a zero--order quantity in a perturbative framework, $\propto a_\CD^{-3}$, (ii) the constant--curvature,
a first--order quantity (if a flat background was perturbed),  $\propto a_\CD^{-2}$, and (iii) the backreaction terms as second--order quantities 
$\propto a_\CD^{-1}$  
feature decay--rates that compensate the differences in their initial conditions magnitudes, if the volume scale factor is assumed
to evolve until $a_\CD (z=0) \approx 1000$ \cite{lischwarz}. 

\begin{figure}[htbp]
\begin{center}
\includegraphics[width=7cm,height=8cm]{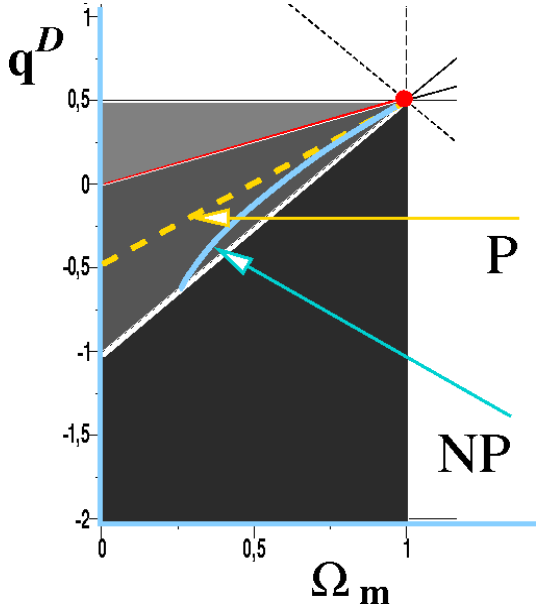}
\caption{\label{fig:zoom}
The unstable sector in Figure~\ref{fig:cosmicstates} that expresses the global instability of the standard model is shown together with the scaling behavior of the leading perturbative mode $\rm P$ discussed in this and the last
subsections. Again, the volume deceleration parameter $q^\CD$ is plotted against the effective density parameter $\Omega_m^\CD$.
This scaling mode (corresponding to the coupling parameter $r= -1/5$ for the scaling index $n= -1$) is shown as a dashed line. It originates from the Einstein--de Sitter model in the center and
ends on the curvature--dominated attractor $q^\CD = 2r/(1+r) =  -1/2$. This scaling solution lies in the quintessence regime, defined by the mean field
description of a morphon field. Recall that it lies in between the line ending at $q^\CD = 0$ (models with 
Friedmannian kinematics with constant negative curvature) and the line ending at $q^\CD = -1$ (a morphon modeling a cosmological constant).
The indicated line $\rm NP$ expresses our expectation of a non--perturbative, non--scaling
solution that would fully explain Dark Energy today, while starting in the vicinity of the Einstein--de Sitter model.}
\end{center}
\end{figure}
\vspace{-10pt}
Thus, the expectation is that a non--perturbative treatment, allowing for an evolving background, would confirm our
extrapolation of the perturbative mode and would even produce a further weakening of the decay rates of the backreaction
terms, eventually coming closer to the behavior of a bare cosmological constant, as speculated in Figure~\ref{fig:zoom}.  
Note that such a behavior, or the more extreme case of a growing backreaction term corresponding to a phantom quintessence in the scalar field
correspondence, must be understood on the grounds that we are looking at the fluctuations within a negatively curved space section. In the course of 
evolution of the averaged scalar curvature, we know that the backreaction mechanism draws 'potential energy' from curvature, and converts
it into an excess of `kinetic energy' that implies the observed weakening of the decay of fluctuations. It is therefore misleading to think about fluctuations as
evolving on a fixed background, i.e. in `Newtonian terms'.
In this context it is worth recalling that, if the employed perturbative framework is `quasi--Newtonian', then this also implies that backreaction terms appear as surface terms \cite{russetal}, \cite{kolbetal}, \cite{lischwarz}, demonstrating that 
we are not describing fluctuations in a curved Riemannian space section in which case the principal scalar invariants of
extrinsic curvature fluctuations cannot be represented through surface terms (compare Subsect.~\ref{subsect:newtonian}).

The fact that already a perturbative mode entails departures of the averaged model from the standard model 
(a `global' instability) means that the
architecture of current N--body simulations and its determining parameters of the concordance cosmology is challenged and it might be
overrestricted for the correct description of the Late Universe: a (possibly indirect) impact of a few percent would already have
severe implications for the demand of `high--precision' cosmology.
This statement needs consolidation in terms of quantitative considerations, an issue
that is very involved and, at present, not conclusive. We shall just add a few remarks
below.

\subsubsection{A few words on quantitative estimates of backreaction}
\label{subsubsect:estimates}

Based on the above--discussed scaling behavior of backreaction that is suggested by perturbation theory, we may discuss
typical magnitudes of backreaction that are expected to be reached today. Since such estimates strongly rely on an extrapolation of a perturbative mode, they are
merely indicative, but they give us an intuition of where we stand with perturbative calculations. 

First, if we naively (i.e. without investigating a sensible re--interpretation of observational data within the new
framework) track the perturbative scaling solution from standard Cold Dark Matter initial data on `some large scale' of
the order of the observable Universe, then the  
comparison of (\ref{om1}) with (\ref{om2}) shows that backreaction is expected to fall short by a large amount to fully explain Dark Energy,
e.g. setting $\CQ_{\initial\CD} = \Lambda$ we obtain with $a_{\now\CD} \approx 1000$, 
$-4\cdot \Omega^{\now\CD}_{\CQ^S}=2 \cdot 10^{-3}\Omega^{\now\CD}_{\Lambda} \approx 0,0015$, which still lies close to the perturbative regime.
The initial data taken assume that the intial expansion fluctuation amplitude is independent and does not necessarily derive from density fluctuations.
Estimates in the literature range from values (perturbative) of $0.004$ for an inhomogeneity--induced $\Lambda-$parameter \cite{vanderveldetal} 
up to  $\Omega^{\now\CD}_{\CQ} \approx -0.05 \cdots -0.26$ (Lyman-$\alpha$ absorbers in the redshift range $z \in \lbrace 3.8 , 2\rbrace$\cite{rasanen}, which may at best be taken as an indication of a discrepancy between perturbative model estimates and the way of how we interpret observational data. 

Second, if we look at those estimates in a scale--dependent way, i.e. taking into account that the influence of backreaction
must be compared to $\Lambda$ on the observational scales at which we postulate a Dark Energy component, then
the answer is more sensible: taking initial data for a standard Cold Dark Matter model from \cite{bks} and translating the 
effect on the time--history of $\Omega^{\CD}_{\CQ}$ into the relativistic context, we would start to explain
the value of $\Omega^{\now\CD}_{\Lambda}$ by the perturbative scaling mode today on scales of typically {\em below} 100 Mpc, if that region is at 2--$\sigma$ variance level in the initial conditions. For a typical such region (at 1--$\sigma$) we
would not compensate $\Lambda$, but would talk about a significant effect in magnitude.

The number of pitfalls in the above considerations is, however, large. A re--interpretation of the other cosmological parameters
in terms of their scale--depen\-dence is mandatory, especially since the indirect influence of a non--vanishing backreaction
on the other cosmological parameters has been found to be crucial and actually is expected to {\em largely} outweigh the magnitude effect
in $\Omega^{\CD}_{\CQ}$ (compare the discussion in \cite{bks}). Therefore, it might not be a good idea to judge the influence of
backreaction based on the magnitude of $\Omega^{\CD}_{\CQ}$ itself. We have to investigate realistic models beyond perturbation theory at a 
fixed background, before we can reliably discuss quantitative estimates from models.
 
\subsection{Issues of interpretation of backreaction within observational cosmology}

\subsubsection{A first step: a quasi--Friedmannian template metric}
\label{subsubsect:mockmetric}

The particular form of the metric for an effective approximation of the inhomogeneous
Universe that springs to mind has been suggested and thoroughly discussed by Paranjape and Singh \cite{singh2}, who consider the metric form 
\begin{equation}
{}^4 {\bf g}^\CD \;=\; -dt^2 + a_\CD^2 \gamma^\CD_{ij}\,dx^i \otimes dx^j \;\;,
\end{equation}
with the {\em volume scale factor} $a_\CD (t)$ on a mass--preserving compact domain $\CD$ that is specified in terms of the exact kinematical equations, and a 
(domain--dependent) {\em effective constant curvature three--metric} with coefficients 
$\gamma^\CD_{ij}$ that, as opposed to \cite{singh2},
may also allow for a time--parametrization of the constant--curvature 
appearing in $\gamma^\CD_{ij}$.
The concrete form of the 3--metric coefficients we consider reads:
\begin{equation}
\label{tempmetric}
 \gamma^\CD_{ij}\;=\;
\left(\frac{dr^2}{1-\kappa_{\CD}(t)r^2}+d\Omega^{2}\right)\;\;,
\end{equation} 
where $\kappa_\CD (t)$ corresponds to the (domain--dependent) constant curvature of the template
space at time $t$, and $d\Omega^{2}=r^2(d\phi^{2}+\sin^{2}(\phi)d\psi^2)$.

It should be emphasized that this template metric must not be a dust solution of Einstein's equations
\cite{boundsoncurvature}, \cite{rasanenk(t)}
(the effective fluid of an averaged dust model also features a geometrical pressure).

The reason why we wish to allow for an explicit time--dependence of 
the `curvature constant' $\kappa_{\CD}$ is given
by the key--insight that the constant--curvature evolution is not identical with that of the averaged 
3--Ricci curvature of an inhomogeneous universe model, if the degrees of freedom in
inhomogeneities (kinematical backreaction) are taken into account, e.g. \cite{buchert:grgdust},
\cite{buchert:static}, \cite{rasanen}. This effective metric provides an alternative dynamical picture to the
thoughts recently advanced by Kasai \cite{kasaiSN}, who investigated the goodness of fit to supernova data
for Friedmannian models without cosmological constant, but {\em different} curvature
parameters. Thus, while a {\em single} standard model without cosmological constant cannot
account for the supernova data, two such models -- if applied to low-- and high--redshift
data separately -- would \cite{kasaiSN}. In \cite{SN1a} we are currently investigating this model for the
purpose of fitting supernova data. This fit must be constrained by CMB observations, since otherwise we could
not significantly distinguish the curvature evolution with backreaction from the constant--curvature evolution in
a narrow range of redshifts \cite{franca}, \cite{ichikawa}, \cite{bruceSN}.

This form of an effective metric can be motivated on the grounds that Ricci flow renormalization of the 
average characteristics on a bumpy geometry, {\it cf.} Subsect.~\ref{subsubsect:ricciflow}, would produce a constant--curvature slice, 
{\em but} only at a given instant of time. In general, such a flow has singularities, if the Ricci tensor is non--positive, and a constant--curvature model is reached only after
subsequent steps of surgery of the manifold. However, if we assume {\em intrinsic curvature fluctuations} (not the averaged curvature), i.e. terms
like $\average{(\CR - \average{\CR})^2}$ to be subdominant over kinematical (extrinsic curvature) fluctuations,
then we may assume that the actual inhomogeneous metric (at one instant of time!) is already close to a constant--curvature metric, in which case Ricci flow smoothing
may be free of singularities. In any case, the disclaimer of using such a simple metric for e.g. calculating luminosity distances is still that we neglect the effect of inhomogeneities on light propagation. This issue we address now.

\subsubsection{Averaging on the lightcone}
\label{lightcone}

Here, the most important step that would considerably advance the management of observational data, 
will be to investigate the averaging formalism on the lightcone.
Such a framework is currently being constructed\cite{lightcone}. It relates not only to all aspects of observations
in terms of distances within inhomogeneous cosmologies, but also links directly to initial
data in the form of, e.g., CMB fluctuation amplitudes and the integrated Sachs--Wolfe effect. Relating lightcone averages to cosmological model averages is also possible and is in the focus
of this investigation. For example, a closed smooth lightfront would enclose a region
of space that is characterized by the evolution of the volume scale factor employed in
this report. The consequences of a quantitative importance of an integrated backreaction history, described through a propagating morphon along the lightcone, are obvious. Applying generic redshift--distance
relations e.g. to galaxy surveys would put us in the position to better understand the
actual distribution of galaxies that are currently mapped with the help of FLRW distances.  
If expansion fluctuations are a dominant player on large--scales, we can imagine that also
the galaxy density maps would be affected.
This attempt is non--perturbative in the sense that the fully nonlinear optical propagation equations are averaged; quasi--Newtonian estimates
may capture (on the background--defined lightcone) localized perturbation magnitudes \cite{vanderveldetal}, but they suffer from the same restrictions
as those discussed in Subsect.~\ref{subsubsect:quasinewtonian}, i.e. the averaged curvature of the lightcone 
integrated over its full propagation history may substantially deviate from a perturbed background--defined curvature. 
(Compare here also the remarks on metrical properties of spacetime at the end of the following subsection.)

\subsubsection{Direct measurement of kinematical backreaction}
\label{subsubsect:measurement}

If we ask whether the kinematical backreaction term ${\CQ}_\CD$ is observable, the answer within
a Newtonian (or quasi--Newtonian) framework is straightforward: on the observable domain $\CD$, ${\CQ}_\CD$ is built from invariants
of the peculiar--velocity gradient in a Newtonian model. Ignoring geometrical fluctuations on regional scales may not be
unrealistic to estimate this term from high--resolution maps of peculiar--velocities. More precisely, we need to carefully map the
gradient of the peculiar--velocity to build the Newtonian approximation of ${\CQ}_\CD$.
We so have to ignore the fact that in a relativistic setting ${\CQ}_\CD$ cannot be 
represented through invariants of a gradient, which is derived from a vector field. Existing catalogues are,
however, too small and usually, for the definition of peculiar--velocities, the prior of a Friedmannian model is imposed, which therefore would only return the cosmic variance around the assumed Friedmannian background in a likely untypical patch of the Universe
that is statistically affected by boundary conditions \cite{zehavietal}, \cite{giovanellietal}.
The measurement of ${\CQ}_\CD$ on small scales may also provide a negative value, i.e. irrelevant for a direct
large--scale estimate of Dark Energy, but relevant for a scale--dependent evaluation of ${\CQ}_\CD$.  Indications for a
shear--dominated ${\CQ}_\CD$ on scales of about 100Mpc were discussed in the Newtonian analysis \cite{bks}.  
Two papers are of particular interest here: by taking the sampling anisotropies of the velocity field explicitly into account,
Reg\"os and Szalay \cite{regoesszalay}, already in 1989, reported a large effect (40 \%) of the dipole and quadrupole anisotropies 
on the estimated bulk flow of an elliptical galaxy sample; around the same time, using the Eulerian linear approximation, G\'orski \cite{gorski}
already showed that the velocity field is significantly correlated even on scales of 100 Mpc.
The measurement of the shear field related to weak gravitational lensing can add further information for backreaction on regional scales \cite{schimd}.
On large scales, on the other hand, we know several observational data that could place constraints
on the value of kinematical variables \cite{ellis:dahlem}.  
`Global' bounds on ${\CQ}_\CD$, where $\CD$ is of the order of the
CMB scale, can be inferred from work of Maartens et al.
\cite{maartensetal1}, \cite{maartensetal2}.

In this context, the question 
whether and how close our observers have to be at the center of a 
regional `Hubble bubble', that probes the expected negative curvature region for positive backreaction,
furnishes relevant observational input \cite{tomita1}, \cite{tomita2}, \cite{tomita3} \cite{linde2}, \cite{LTBgron}. The scale of this `reduced
curvature region'  likely exceeds scales that have been discussed in connection with
peculiar--velocity catalogues.

Another possibility is to exploit the relation of the kinematical backreaction term to Minkowski Functionals, as outlined in
Subsect.~\ref{subsubsect:minkowski}. The problem here is to identify the boundary of the averaging region with a surface of constant
peculiar--velocity potential. Again we need peculiar--velocity data or, alternatively, a model--dependent relation between iso--density and 
velocity potential surfaces; the relativistic geometrical effects are again ignored.
The boundary of the averaging region plays a crucial role, since it carries higher--order correlations of the velocity 
distribution encoding the history of structure formation, and hence the backreaction history that was identified as the 
source of the general expansion law (\ref{generalexpansionlaw}). Measuring Minkowski Functionals of iso--velocity potential
surfaces thus directly mirrors the fact that  ${\CQ}_\CD$ is determined through all orders of the correlation functions.
In this line it is important to point out that, even if the fluctuations in number density (the first moment of the 
galaxy distribution) and in the two--point correlation function (or the power spectrum, i.e. the second moment) may not be significant, fluctuations may show up especially in higher moments, since those determine the morphology of the averaging
region (the phase correlations). An investigation of subsets from the IRAS catalogues revealed large morphological
fluctuations up to scales of 200 Mpc that are significant on scales of the order of several tens of Megaparsecs, while on
the scale around 10 Mpc these fluctuations disappeared \cite{kerscher:mf}, \cite{kerscher:pscz}. This has been confirmed by a
recent analysis of SDSS data \cite{SDSS}, although here deviations were not so dramatic, an issue that has to be (and is currently) addressed with the help of  a substantially improved data set.     

A direct determination of metrical properties of spacetime rather than properties of the matter distribution from observational data furnishes a promising programme that relates to all the issues outlined here \cite{hellaby:mass}, \cite{luhellaby}.
This programme relates to the fully relativistic considerations pursued here as opposed to the prior of a quasi--Newtonian model that usually enters
into the interpretation process. Here it is important to realize that, irrespective of the small {\em magnitude} of the field strength in a weak--field
situation, its derivatives may be important.
If we consider space to be Euclidean and the gravitational field of the mass distribution to be a quasi--Newtonian perturbation, then we 
may not correctly characterize the effect of intrinsic curvature that is built in a highly nonlinear way from derivatives of the metric tensor. 
There are effects due to the morphological properties of the gravitational field, e.g. the volume effect being the simplest morphological characteristic mentioned in Subsect.~\ref{subsubsect:ricciflow}. As
Hellaby \cite{hellaby:volume} showed, a volume matching of a Friedmannian template model to such a distribution implies an error of 10--30 percent which may be interpreted as a volume effect in a mass--preserving smoothing procedure due to a factor of the order of  $\pi^2 /6$ with which the Euclidean volume and the Riemannian volume of a ball differ \cite{buchertcarfora3}. Such a factor cannot be regarded as a perturbation of $1$. Otherwise stated:
the metrical properties of space could be very different
from Euclidean in terms of the morphology (volume, shape, connectivity) of the gravitational field, not in terms of its magnitude.

\subsubsection{A common origin of Dark Energy and Dark Matter?}
\label{subsubsect:dark}

Several times we have already pointed out that the scale--dependence of observables
is key to understand the cosmological parameters in the present framework.
Viewing observational data with this additional discrimination power of a scale--dependent interpretation of backreaction effects, there is furthermore a link to the {\em Dark Matter problem} that certainly is important to be understood in relation to
sources, i.e. Dark Matter particles, but there is also a kinematical contribution that may alter existing strategies of Dark Matter search.

Concentrating on the Dark Energy problem has led us to focussing on a positive contribution
of ${\CQ}_\CD$ on large scales. However, as already mentioned above in the context of peculiar--velocity catalogues, the 
kinematical backreaction ${\CQ}_\CD$ itself can also be negative, and a sign--change may actually happen by going to smaller scales. Looking at the phenomenology
of large--scale structure reveals strongly anisotropic patterns, so that it is not implausible
that on the scales of superclusters of galaxies we would find\footnote{This was actually found in the Newtonian investigation \cite{bks} that, however, suffers
from the fact that ${\CQ}_\CD$ is restricted to drop to zero on the periodicity scale of the fluctuations.}
a shear--dominated ${\CQ}_\CD < 0$. Thus, again as a result of its scale--dependence, the kinematical backreaction parameter
can potentially be the origin of {\em Kinematical Dark Energy}, but also of 
{\em Kinematical Dark Matter} \cite{buchert:valencia}.

Mapping kinematical backreaction with a `morphon field' opens further links to previous
studies that tried to model Dark Energy {\em and} Dark Matter by a scalar field
(\cite{paddy:darkmatter}, \cite{arbey:complex} and references to earlier work therein).
Other explanations to unify the description of Dark Energy and that of Dark Matter may also be put into perspective \cite{fuzfaalimi}.
With this in mind, the volume deceleration functional (\ref{deceleration})
can change sign too, but this crucially depends on the value of the matter density parameter $\Omega^\CD_m$.
We infer from Eq.~(\ref{deceleration}) that,
for a small value of $\Omega^\CD_m$, a smaller negative value of
$\Omega^\CD_{\CQ}$ is needed to obtain volume acceleration, $q^\CD < 0$. 
Since this problem touches on a scale--dependent understanding of cosmological parameters, we now propose
a strategy to properly address this issue.

\subsubsection{Multi--scale analysis of backreaction}
\label{subsubsect:multiscales}

Let us discriminate different spatial scales by a suitable partitioning of space sections. We denote by $L_\CH$  a scale larger
than the homogeneity scale, say the Hubble--scale,
by $L_\CE$ the scale of a typical void, and by $L_\CM$ a typical scale of a matter--dominated region
(e.g. galaxy clusters) \cite{buchertcarfora:Q}. 
In standard cosmology we would require $\Omega^{\now\CH}_m \approx 1/4$ including Dark Matter. Hence, in order to find
volume acceleration {\em today}, {\it cf.} Eq.~(\ref{accelerationcondition}), we would need $-\Omega^{\now\CH}_\CQ > 1/16$.
If, however, the global value of the matter parameter on the scale $L_\CH$ is smaller, then also the needed amount of 
backreaction in a Hubble--domain $\CH$ is smaller.
Now, we discuss that it is indeed the case that the matter density parameter drops substantially at around the scale 
$L_\CE$ in a cosmological slice that is volume--dominated by voids.

We employ the averaged Hamiltonian constraint (\ref{averagedhamilton}), 
and assume that  
a domain as large as $\CH$ is formed out of  
a union of underdense regions ${\CE}$ and a union of occupied overdense 
regions ${\CM}$.
We further consider the following picture that complies with what we see in the present--day Universe: we require the 
volume Hubble expansion to be subdominant in matter--dominated regions and,
on the other hand,  the averaged density to be subdominant in devoid regions.
In the first case, an expansion or contraction would negatively contribute and so would, e.g., 
enhance a negative
averaged curvature, in the second case, the presence of a low averaged density would positively contribute.
We can therefore reasonably expect that the following idealization of the distributions would not substantially impair
the overall argument: we model voids with $\langle\varrho\rangle_\CE = 0$ and
matter--dominated regions with $H_\CM = 0$ (corresponding to the stable clustering hypothesis).  
We also introduce a parameter for the occupied volume fraction,
$\lambda_\CM := V_\CM / V_\CH$, where $V_\CM$ denotes the total volume of the union
of occupied regions $\CM$, that 
may be chosen more conservatively to weaken this idealization.
Thus, 
we would have:
\begin{equation}
\label{averagehamiltonE}
\averageE{\CR}\;=\;-6H_\CE^2  -{\CQ}_\CE +2\Lambda\;\;\;;\;\;\;
{\averageM{\CR}\;=\;-{\CQ}_\CM} + {16\pi G}\langle\varrho\rangle_\CM +2\Lambda\;,
\end{equation}
together with\footnote{The fact that we expect the global Hubble parameter to be slightly smaller than the one measured on
the scale of voids could be used, of course with more refined assumptions, to observationally determine the volume fraction
$\lambda_\CM$. It will be these refined assumptions together with a scale--dependent treatment of other relevant variables that put
us in the position to seriously think about an observational determination of the void volume fraction that is certainly one of the key--parameters
of a scale--dependent cosmology.} $\quad H_\CH = (1-\lambda_\CM ) H_\CE \quad{\rm and}\quad \langle\varrho\rangle_\CH =
\lambda_\CM \langle\varrho\rangle_\CM$.

Consider for the moment the case where
the kinematical backreaction terms in the above 
equations are negligible and that there is no cosmological constant. Then, we infer that the averaged scalar curvature must be {\em negative} on 
domains $\CE$ and {\em positive} on domains $\CM$, what obviously complies with what we expect.
We form the `global' cosmological parameters by dividing by $H^{2}_{\CH}$, `regional' cosmological paramters may be
introduced by dividing by $H^{2}_{\CE}$, if we wish to relate sources to the regionally measured Hubble parameter.
The introduction of cosmological parameters on the scale $L_\CM$ is pathological and useless. 
With our assumptions the matter density parameter
$\Omega^\CH_m$ can be traced back to the average density in matter--dominated
regions, $\langle\varrho\rangle_\CH \cong \lambda_\CM \langle\varrho\rangle_\CM$,
and thus, the global density parameter can be reconstructed out of an observed $\langle\varrho\rangle_\CM$ on the scale $L_\CM$.
Therefore, we find a smaller value for the density parameter on the global scale, depending on the value of the volume fraction of occupied regions, 
as a consequence of the compensation (through conservation of the total mass) of the missing matter in the regions $\CE$.

The volume fraction is a sensible quantity since it depends on the coarsening of the distribution.
We know that even in matter--dominated regions $\CM$ the matter distribution in luminous matter is very spiky  leaving a lot of volume
to empty space. Whether this argument carries over to all matter depends on how smoothly Dark Matter is distributed.
In relativistic cosmology it is crucial that, unlike for the mass, {\em there is no equipartition of curvature} in Riemannian space sections (there is more volume available in negatively curved regions than in positive ones); therefore, Newtonian estimates always provide a conservative upper limit on a realistic volume fraction. 
It is not implausible that a realistic value for $\lambda_{\now\CM}$ could be much smaller than anticipated by Newtonian simulations
that employ a fairly large coarsening scale (\cite{colberg}; other estimates give a larger value for the void volume fraction, see discussion and references
in \cite{rasanen}, \cite{buchertcarfora:Q}).

Finally, it should be noted that a scale--dependent analysis may be performed for a given slicing of spacetime, as above, but
we may also expose the particular situation of observers, who perform measurements in matter--dominated regions,
to a refined analysis of a scale--dependent slicing. Such a picture has been recently advanced by Wiltshire and coworkers \cite{wiltshire07}, \cite{wiltshire_exact}, \cite{wiltshire_obs},
distinguishing cosmic  from the observer's time,
and this would involve considerations of spatial renormalization of average characteristics that we briefly discussed
in Subsect.~\ref{subsubsect:ricciflow}.

\subsection{A short conclusion: opening {\em Pandora's Jar}}

Let us conclude by stressing the 
most important issue: {\em quantitative relevance of backreaction effects}. 
Even if all these efforts would `only' nail down an effect of a few percent, rather than $75$ percent, these studies
would have justified their quantitative importance for observational cosmology, and what is
to be expected, would substantially improve our understanding of the Universe. 

Especially the recent efforts, spent on the backreaction problem by a fairly large number of researchers,
added substantial {\em qualitative} understanding to the numerous previous efforts that were 
undertaken since George Ellis initiated this discussion in 1984 \cite{ellis} (see references in \cite{ellisbuchert}).
The issue remains unresolved to date: an explanation of Dark Energy along these lines
is attractive, not only because it naturally explains the coincidence problem. From what has been said, it is also physically plausible, but a reliable and unambiguous estimate of the actual
influence of these effects is lacking. This situation may change soon and for this to happen it requires
considerable efforts, for which some possible strategies have been outlined in this section.  
 
After those results are coming in, we may face a more challenging situation than anticipated
by the qualitative understanding that we have. For example, while the explanation of Dark Energy by quintessence (or phantom quintessence) still allows to hide the physical
consequences behind a scalar field that is open for a number of explanations, the mapping
of a scalar field to the backreaction problem, as in the mean field description outlined in Subsect.~\ref{subsect:morphon}, 
can no longer keep a phenomenological status: {\em fluctuations exist and can be measured}.
There are no free parameters, there are initial data that can be constrained.

Despite being premature, let us speculate that the outcome is
i) a confirmation of the qualitative picture of a backreaction--driven cosmology, but ii) a quantitative problem to reconcile
this picture with the data in the sense that there is not enough time for the mechanism
to be sufficient. In that situation we `lost' the standard model for a correct description of the
Late Universe, and we do not reach a full explanation of Dark Energy -- unless -- we allow for initial data that
are non--standard. This situation would in turn ask for a comprehensive understanding of these required
initial data, hence reconsideration of inhomogeneous inflationary models \cite{inhominflation} and their
fluctuation spectrum at the exit epoch. 
As further discussed in \cite{buchert:static}, globally inhomogeneous initial data may arise by the very same mechanism: 
if backreaction plays a role due to the generic coupling of 
fluctuations to intrinsic curvature in the Late Universe, then this coupling may have been efficient also in the Early Universe.
Is it conceivable that the Universe evolved out of a spaceform with strongly positive averaged scalar
curvature that, during inflation, acquires `flatness' on average, but at the end leaves an imprint in the fluctuation spectrum as a remnant of the kinematical conversion of curvature energy? We opened Pandora's Jar.

Notwithstanding, I would consider such a situation as the beginning of a fruitful
development of cosmology. As previously mentioned, the issues of scale--dependence
of observables, the priors underlying interpretations of observations, the large subject
of Dark Matter and, of course, the issue of Dark Energy, will be all interlocked and ask for a comprehensive realistic treatment
beyond crude idealizations.  
\vspace{-7pt}
\begin{acknowledgements}
I acknowledge support and hospitality by Observatoire de Paris and Universit\'e Paris 7.
Special thanks go to Henk van Elst and Syksy R\"as\"anen for discussions on a preliminary
draft, and to Ruth Durrer for a careful reading of the final manuscript with fruitful suggestions, 
and for discussions during a working visit to Geneva University whose support is also greatfully
acknowledged.
This report is an extension of lecture notes \cite{buchert:mangaratiba} that include further 
acknowledgments.

\end{acknowledgements}

\newpage

\renewcommand{\theequation}{A.\arabic{equation}}
\setcounter{equation}{0}  
\section*{Appendix: Averaged ADM equations for non--vanishing lapse function}

For completeness, we here add the general Einstein equations for a specified foliation of spacetime employing lapse and shift functions
according to the {\it Arnowitt--Deser--Misner}, short {\it ADM formulation} \cite{adm1962}, \cite{smarryork}, and discuss the resulting system of
spatially averaged equations for vanishing shift.

\subsection*{\it The ADM equations recalled \footnote{Notation: a semicolon denotes covariant derivative with respect to the 4--metric
with signature $(-,+,+,+)$ (the units are such that $c=1$), a double vertical slash
covariant spatial differentiation with respect to the 3--metric, and 
a single slash denotes partial differentiation with respect to the local coordinates; greek indices run through $0 \dots 3$, and latin indices through
$1 \dots 3$; summation over repeated indices is understood.}}

Let $n_{\mu}$ be the future directed unit normal to a three--dimensional Riemannian hypersurface $\Sigma$.
The projector into $\Sigma$, $h_{\mu\nu} = g_{\mu\nu} + n_{\mu}n_{\nu}\;,\;\
(\Rightarrow h_{\mu\nu}n^{\mu} = 0\;,\;h^{\mu}_{\;\,\nu}h^{\nu}_{\;\,\gamma}=
h^{\mu}_{\;\,\gamma})$, induces in $\Sigma$ the 3--metric 
\begin{equation}
h_{ij}: = g_{\mu\nu}h^{\mu}_{\;\,i}h^{\nu}_{\;\,j}\;\;\;.
\end{equation}
Let us write
\begin{equation}
n_{\mu} = N (-1,0,0,0)\;\;\;\;,\;\;\;\;n^{\mu} = \frac{1}{N}(1,-N^i )
\;\;,
\end{equation}
with the {\em lapse function} $N$ and the {\em shift vector} components $N^i$. Note that 
$N$ and $N^i$ determine our choice of coordinates.

\noindent
From $n_{\mu} = g_{\mu\nu}n^{\nu}$ we find $g_{00}=-(N^2 - N_i N^i )$; $g_{0i}=N_i$; $g_{ij}=h_{ij}$ and,
using local coordinates $x^i$ in a $t=const.$ hypersurface $\Sigma$ with $3-$metric $g_{ij}$, 
setting $x^0 = t$ and $dx^0 = dt$, the line element becomes:
\begin{eqnarray}
ds^2 = - (N^2 - N_i N^i )\,dt^2 + 2N_i  \,dx^i dt + g_{ij}\,dx^i \otimes dx^j \;\;\nonumber\\
= -N^2 dt^2 + g_{ij}\, (dx^i + N^i dt) \otimes (dx^j + N^j dt) \;\;.\quad
\end{eqnarray}
Introducing the extrinsic curvature on $\Sigma$ by
\begin{equation}
K_{ij} : = -n_{\mu;\nu}h^{\mu}_{\;\,i}h^{\nu}_{\;\,j} = -n_{i;j}\;\;,
\end{equation}
we obtain the  {\it Arnowitt--Deser--Misner}, short {\it ADM equations} \cite{adm1962}, \cite{smarryork}, \cite{ellisvanelst:cargese}:

\smallskip

\noindent
{\it Energy (Hamiltonian) constraint:}
\begin{equation}
\label{hamiltonN}
{\CR} - K^i_{\;\,j}K^j_{\;\,i} + K^2 = 16\pi G \varepsilon + 2\Lambda
\;\;\;,\;\;\;
\varepsilon: = T_{\mu\nu}n^{\mu}n^{\nu}\;\;\;;
\end{equation}

\noindent
{\it Momentum (Codazzi) constraints:}
\begin{equation}
K^i_{\;\,j||i} - K_{||j} = 8\pi G J_j \;\;\;,\;\;\;
J_i : = - T_{\mu\nu}n^{\mu}h^{\nu}_{\;\,i}\;\;\;;
\end{equation}

\noindent
{\it Evolution equation for the first fundamental form:}
\begin{equation}
\frac{1}{N}\frac{\partial}{\partial t} g_{ij} = -2 K_{ij} + 
\frac{1}{N} (N_{i||j} + N_{j||i}) 
\;\;\;;
\end{equation}

\noindent
{\it Evolution equation for the second fundamental form:}
\begin{equation}
\frac{1}{N} \frac{\partial}{\partial t} K^i_{\;\,j} = {\CR}^i_{\;\,j} + 
K K^i_{\;\,j} - \delta^i_{\;\,j} \Lambda 
- \frac{1}{N}{N^{||i}}_{||j} + \frac{1}{N} \left(
K^i_{\;\,k}N^k_{\;\,||j} - K^k_{\;\,j}N^i_{\;\,||k} + N^k K^i_{\;\,j||k}
\right)\nonumber
\end{equation}
\begin{equation}
- 8\pi G  ( {S}^i_{\;\,j} + \frac{1}{2}\delta^i_{\;\,j} 
(\varepsilon - {S}^k_{\;\,k}))\;\;\;\;;\;\;\;\;
{S}_{ij} : \,=\, T_{\mu\nu} h^{\mu}_{\;\,i}h^{\nu}_{\;\,j}\;\;.
\end{equation}

\vspace{5pt}

\noindent
For the trace parts of (A2c) and (A2d) we have:

\smallskip

\begin{equation}
\frac{1}{N} \frac{\partial}{\partial t} g = 2g (-K + \frac{1}{N}N^k_{\;\,||k})
\;\;\;,\;\;\;g:=\det (g_{ij} )\;\;\;;
\end{equation}
\begin{equation}
\label{traceK}
\frac{1}{N} \frac{\partial}{\partial t} K =  {\CR} + K^2 - 4\pi G (3\varepsilon - 
{S}^k_{\;\,k} ) - 3\Lambda  
- \frac{1}{N}{N^{||k}}_{||k} + \frac{1}{N} N^k K_{||k}\;\;\;. 
\end{equation}

\vspace{10pt}

In relativistic cosmology it is often assumed that  the energy--momentum tensor has the form
of a perfect fluid $T_{\mu\nu} = 
\varepsilon u_{\mu}u_{\nu} + p h_{\mu\nu}$. Also, it is often required that the fluid is
irrotational; putting the shift vector field $N^i = 0$, we then model all inhomogeneities of the 
fluid by the 3--metric and the lapse function.
The lapse function is related to the fluid acceleration in the hypersurface that reduces to the pressure
gradient in {\em fluid--comoving gauge} (see below):  
\begin{equation}
a_i =\frac{N_{||i}}{N}\;\equiv \;\frac{- p_{||i}}{\varepsilon + p}\;\;.
\end{equation}

\smallskip

\noindent
Notice that with this gauge choice the unit normal coincides with the 4--velocity and, especially, the momentum flux
density in $\Sigma$ vanishes. The total time--derivative operator
of a tensor field $\cal F$ along integral curves
of the unit normal, $d/ d\tau\,{\cal F}: = n^{\nu}\partial_{\nu}{\cal F}
= u^{\nu}\partial_{\nu}{\cal F}$ becomes 
\begin{equation}
\frac{d}{d\tau}{\cal F} = \frac{1}{N}\frac{\partial}{\partial t}  {\cal F}\;\;\;,
\end{equation}
since $n^{\nu}{\cal F}_{||\nu} = 0$.
Note that, although the definition of proper time is $\tau: = \int{N dt}$,
the line element cannot be written in the form of the comoving gauge by
measuring `time' through proper time $d\tau = N dt$, 
since $d\tau$ is not an exact form in the case of an inhomogeneous
lapse function. The exterior derivative of the proper
time will involve a non--vanishing shift vector according to the 
space--dependence of the lapse function. Therefore, 
a foliation into hypersurfaces $\tau = {\rm const.}$
with simultaneously requiring $u_{\alpha} = - \partial_{\alpha}\tau$ is not
possible.

\subsection*{\it Averaged ADM equations for vanishing shift}

For vanishing shift vector, as will be our choice for the averaged equations, the line element reads:
\begin{equation}
ds^2 = -N^2 dt^2  + g_{ij}\,dX^i \otimes dX^j \;\;,
\label{comovingslicing}
\end{equation}
where we have written the local coordinates in capital letters now, as in the main text, to indicate that they now label 
comoving fluid elements.

We here recall the results given in \cite{buchert:grgfluid}.
We shall study spatial averages in a hypersurface defined by the choice of the 
in general inhomogeneous lapse function $N$ in the line--element (\ref{comovingslicing}).

We consider perfect fluid sources, i.e. energy density $\varepsilon$ and pressure $p$
with energy momentum tensor 
$T_{\mu\nu} = \varepsilon u_{\mu}u_{\nu} + p h_{\mu\nu}$. Restricting attention to irrotational flows we can, without loss of generality,
write the flow's 4--velocity in the form 
\begin{equation}
\label{4velocity}
u^{\mu} = -\frac{\partial^{\mu}{\cal S}}{h}\;\;;\;\;h=\frac{\varepsilon +p}{\varrho}\;\;,
\end{equation}
together with the decomposition into kinematical parts of the 4--velocity gradient,
\begin{equation}
\label{decomposition}
u_{\mu;\nu} = \frac{1}{3}\Theta h_{\mu\nu} + \sigma_{\mu\nu}  + \omega_{\mu\nu}
- {\dot u}_{\mu}u_{\nu}\;\;,
\end{equation}
where the inhomogeneous normalization of the 4--velocity gradient $h$ 
is given by the injection energy per fluid element and unit restmass,
$d\varepsilon = h d\varrho$ with the restmass density $\varrho$ \cite{israel76};
$\Theta$ is the rate of expansion, $\sigma_{\mu\nu}$ the shear tensor.

The existence of a scalar 4--velocity potential $\cal S$ together with the choice
(\ref{4velocity}) implies that the conservation equations $T^{\mu\nu}_{\;\;\,\,;\nu}=0$
are satisfied, but also that the flow has to be irrotational and that the covariant
spatial gradient of $\cal S$ vanishes
\cite{bruni92a,bruni92b,dunsby}, \cite{buchert:grgfluid}:
\begin{eqnarray}
\label{restriction1}
\omega_{\mu\nu} = h_{\mu}^{\;\,\alpha}h_{\nu}^{\;\,\beta}u_{[\alpha;\beta]} =
- h_{\mu}^{\;\,\alpha}h_{\nu}^{\;\,\beta}\left(\,\partial_{[\alpha}{\cal S}/h\,\right)_{;\beta]}
=0\;\;;\\
\label{restriction2}
{\cal S}_{||\mu} = h_{\mu}^{\;\,\alpha}\partial_{\alpha}{\cal S} = \partial_{\mu}{\cal S}
+ u_{\mu} \dot{\cal S} = 0\;\;,
\end{eqnarray} 
with the covariant time--derivative $\dot {\cal S}:= u^{\mu}\partial_{\mu} {\cal S}\equiv h$.

We now define the averaging operation in terms of Riemannian volume
integration on the hypersurfaces orthogonal to $u^{\mu}$, 
restricting attention to scalar functions $\Psi (t,X^i)$, 
\begin{equation}
\label{average}
\langle \Psi (t, X^i)\rangle_{\cal D}: = 
\frac{1}{V_{\cal D}}\int_{\cal D} \Psi (t, X^i)\,d\alpha_g  \;\;,
\end{equation}
with the Riemannian volume element $d\alpha_g := \sqrt{g} d^3 X$, $g:=\det(g_{ij})$, and 
the volume of an arbitrary compact domain, $V_{\cal D}(t) : = \int_{\cal D} J d^3 X$;
$J:= \sqrt{\det (g_{ij})}$.
We introduce a dimensionless scale factor via the volume (normalized by the volume 
of the initial domain ${V_{\cal D}}_i$):
\begin{equation}
\label{scalefactor}
a_{\cal D} (t) : = \left(\frac{V_{\cal D}}{{V_{\cal D}}_i}\right)^{1/3} \;\;\;.
\end{equation}
This means that we are only interested in the volume dynamics of the domain; 
$a_{\cal D}$ will be a functional of the domain's shape (dictated by the metric) and position. 
We require the domains to follow the flow lines, so that the total restmass 
$M_{\cal D}: = \int_{\cal D} \varrho J d^3 X $
contained in a given domain is conserved. 
Introducing the scaled (t--)expansion ${\tilde\Theta}:=N\Theta$, 
the rate of change of the domain's volume within the spatial hypersurfaces
defines the rate of volume expansion and, through  (\ref{scalefactor}), 
an effective volume Hubble rate: 
\begin{equation}
\langle \tilde\Theta \rangle_{\cal D} = \frac{\partial_t V_{\cal D} (t)}{ V_{\cal D} (t)} = 3 
\frac{\partial_t a_{\cal D}}{a_{\cal D}} =: 3 {H}_{\cal D}\;\;.
\end{equation}
We shall reserve the overdot for the covariant time--derivative
(defined through the 4--velocity $u^{\mu}$):
\begin{equation}
\label{covarianttime}
\frac{\partial}{\partial\tau}:= u^{\mu}\frac{\partial}{\partial \mu}=
\frac{1}{N}\frac{\partial}{\partial t}\;\;,
\end{equation}
and we shall abbreviate the coordinate time--derivative by a prime
in all following equations.
For an arbitrary scalar field $\Upsilon (t,X^i )$ we make essential use of the {\em commutation rule}
\begin{equation}
\label{commutation}
\langle \Upsilon\rangle_{\cal D}' - \langle{\Upsilon}'
\rangle_{\cal D} = \langle \Upsilon\tilde\Theta\rangle_{\cal D} - 
\langle \Upsilon\rangle_{\cal D}\langle\tilde\Theta\rangle_{\cal D}
\;\;,
\end{equation}
or, alternatively, $\langle\Upsilon\rangle_{\cal D}' 
+ 3{H}_{\cal D}\langle \Upsilon\rangle_{\cal D}
= \langle\Upsilon' + \Upsilon{\tilde{\Theta}}\rangle_{\cal D}$.
A simple application is the proof that the total restmass
in a domain is conserved: let $\Upsilon = \varrho$, then 
$\langle \varrho\rangle_{\cal D}'+3 {H}_{\cal  D}
\langle \varrho\rangle_{\cal D} = \langle {\varrho}'+ \varrho\tilde\Theta
\rangle_{\cal D}=0$ according to the local conservation law
$\varrho' + \varrho\tilde\Theta =0$.

\vspace{3pt}

We now consider the scalar parts of Einstein's equations. Their evolution is determined by Raychaudhuri's equation and  
the Hamiltonian constraint (\ref{hamiltonN}). The former can be obtained by inserting (\ref{hamiltonN}) into (\ref{traceK}),
\begin{equation}
\label{raychaudhuri}
{\dot\Theta} = -\frac{1}{3}\Theta^2 - 2\sigma^2 -4\pi G (\varepsilon + 3p) + {\cal A}\;\;, 
\end{equation}
with the rate of shear $\sigma$, $\sigma^2 : = 1/2 \sigma^i_{\;\,j}\sigma^j_{\;\,i}$, and the acceleration divergence ${\cal A}:= ( N^{| k} /N )_{||k}$. Upon averaging these two equations, we can cast the result into a compact form
(to be found under the heading {\it Corollary 2} in \cite{buchert:grgfluid}):
\begin{eqnarray}
\label{effectiveequations}
&3\frac{a_{\cal D}''}{a_{\cal D}} + 
4\pi G \left(\varepsilon_{\rm eff} + 
3p_{\rm eff}\right)\;=\; 0\;\;;\nonumber\\
&6 {H}_{\cal D}^2 - 16\pi G \varepsilon_{\rm eff} 
\;=\; 0\;\;;\nonumber\\
&\varepsilon_{\rm eff}' + 3 {H}_{\cal D}
\left(\varepsilon_{\rm eff} + p_{\rm eff} \right)= 0\;\;,
\end{eqnarray}
with the following fluctuating sources:
\begin{eqnarray}
\label{backreactionsources}
16\pi G \varepsilon_{\rm eff}:= &16\pi G\langle{\tilde\varepsilon}\rangle_{\cal D} 
- {\tilde{\CQ}}_{\cal D}
- \langle{\tilde{\CR}}\rangle_{\cal D}\;\;,\qquad\,\;\quad\nonumber\\
16\pi G p_{\rm eff}:= &16\pi G\langle {\tilde p} \rangle_{\cal D} - {\tilde{\CQ}}_{\cal D} 
+ \frac{1}{3} \langle{\tilde{\CR}}\rangle_{\cal D} - 
\frac{4}{3}{\tilde{\CP}}_{\cal D}\;;
\end{eqnarray}
${\tilde\varepsilon}  := N^2 \varepsilon$ and ${\tilde p}: = N^2 p$ are  
the scaled energy density and pressure of matter, respectively. The 
{\it kinematical backreaction} term is given by:
\begin{equation} 
\label{backreactionQ}
{\tilde{\CQ}}_{\cal D}:= 2 \langle N^2 II\rangle_{\cal D} 
 - \frac{2}{3} \langle N\Theta \rangle^2_{\cal D}\;\;; 
\end{equation}
it is built from the principal scalar invariants 
$2 II: = \Theta^2 -K^i_{\,\;j}K^j_{\,\;i}$ 
and $K^i_{\,\;i} = -\Theta$ of the extrinsic curvature,
\begin{equation}
\label{extrinsiccurvature}
K^i_{\,\;j} = -\frac{1}{2}{g}^{ik}\frac{1}{N}{g}'_{kj}\;\;.
\end{equation}
The averaged scaled scalar curvature and the acceleration backreaction terms read:
\begin{equation} 
\label{backreactionRandP}
\langle{\tilde{\CR}}\rangle_{\cal D}:= \langle N^2{\CR}\rangle_{\cal D}\;\;\;\;;\;\;\;\;
{\tilde{\CP}}_{\cal D}: = \langle {\tilde{\cal A}}\rangle_{\cal D}
+ \Bigl\langle \frac{N'}{N}{\tilde\Theta}\Bigr\rangle_{\cal D} \;\;,
\end{equation}
with the scaled (t--)acceleration divergence ${\tilde{\cal A}}:=N^2 {\cal A} 
= N^2 \left(N^{|i}/N\right)_{||i}$. 

\subsection*{\it Some comments}

With the help of these equations more general matter models can be considered within the kinematically averaged framework.
Notably, scalar field sources and radiation. As for the latter it is interesting that, due to the non--commutativity of averaging and time--evolution,
an averaged radiation cosmos is not described by the familiar law in the homogeneous situation. There are source terms demonstrating that an inhomogeneous radiation cosmos is in an out--of--equilibrium state. An analoguous situation occurs for the {\em dark radiation} part when averaging brane world cosmologies \cite{branes}, where those source terms can be written in terms of effective Tsallis information entropies \cite{hosoya:infoentropy}.
(Note: it is straightforward to interpret the averaged ADM equations for vanishing shift for the choice of a tilted slicing, i.e.
where the $4-$velocity is not required to coincide with the normal on the hypersurfaces: we have to write them for the extrinsic curvature, and not for the expansion tensor, which (up to the sign) agree for our choice.)


%
\end{document}